\documentclass[apj,onecolumn]{emulateapj}
\usepackage{natbib}
\usepackage{lscape,longtable}

\newcommand{\Tew}{\mbox{$T_{\mathrm{ew}}$}}
\newcommand{\Tspec}{\mbox{$T_{\mathrm{spec}}$}}
\newcommand{\chandra}{{\it Chandra}}
\newcommand{\asca}{{\it ASCA}}
\newcommand{\rosat}{{\it ROSAT}}
\newcommand{\xmm}{{\it XMM-Newton}}
\slugcomment{}
\shortauthors{Reese et al.}
\shorttitle{\chandra\ calibration impact on cluster $T_e$}
\begin{document}
%
\title{Impact of \chandra\ calibration uncertainties on galaxy cluster
  temperatures: application to the Hubble Constant}
\author{
  Erik D. Reese\altaffilmark{1,2},
  Hajime Kawahara\altaffilmark{2,3},
  Tetsu Kitayama\altaffilmark{4},
  Naomi Ota\altaffilmark{5}, 
  Shin Sasaki\altaffilmark{3}, and
  Yasushi Suto\altaffilmark{2,6,7}
}
\altaffiltext{1}{Department of Physics and Astronomy, University of
  Pennsylvania, 209 South 33rd Street, Philadelphia, PA, 19104, USA}
\altaffiltext{2}{Department of Physics, The University of Tokyo, 
  Tokyo 113-0033, Japan}
\altaffiltext{3}{Department of Physics, Tokyo Metropolitan University,
  Hachioji, Tokyo 192-0397, Japan}
\altaffiltext{4}{Department of Physics, Toho University,  Funabashi,
  Chiba 274-8510, Japan}
\altaffiltext{5}{Department of Physics, Tokyo University of Science
  1-3 Kagurazaka, Shinjuku, Tokyo 162-8601, Japan}
\altaffiltext{6}{Research Center for the Early Universe, Graduate School
  of Sciences, The University of Tokyo, Tokyo 113-0033, Japan}
\altaffiltext{7}{Department of Astrophysical Sciences, 
  Princeton University, Princeton, NJ 08544, USA}
\email{erreese@physics.upenn.edu}
\begin{abstract}
We perform a uniform, systematic X-ray spectroscopic analysis of a
sample of 38 galaxy clusters with three different
\chandra\ calibrations.  The temperatures change systematically
between calibrations.  Cluster temperatures change on average by
roughly $\sim 6$\% for the smallest changes and roughly $\sim 13$\%
for the more extreme changes between calibrations.  We explore the
effects of the \chandra\ calibration on cluster spectral properties
and the implications on Sunyaev-Zel'dovich effect (SZE) and X-ray
determinations of the Hubble constant.  The Hubble parameter changes
by $+$10\% and $-$13\% between the current calibration and two
previous \chandra\ calibrations, indicating that changes in the
cluster temperature basically explain the entire change in $H_0$.
Although this work focuses on the difference in spectral properties
and resultant Hubble parameters between the calibrations, it is
intriguing to note that the newer calibrations favor a lower value of
the Hubble constant, $H_0 \sim 60$ km s$^{-1}$ Mpc$^{-1}$, typical of
results from SZE/X-ray distances.  Both galaxy clusters themselves and
the details of the instruments must be known precisely to enable
reliable precision cosmology with clusters, which will be feasible
with combined efforts from ongoing observations and planned missions
and observatories covering a wide range of wavelengths.
\end{abstract}

\section{Introduction} 
\label{sec:intro}

Galaxy clusters have played a major role in determining cosmological
parameters.  Abundances of galaxy clusters have placed useful
constraints on the fluctuation amplitude, $\sigma_8$, and the matter
density of the universe, $\Omega_M$ \citep[e.g.,][]{henry1991,
  viana1996, bahcall1997, eke1998, borgani2001, reiprich2002,
  schuecker2003, henry2004, mantz2008, vikhlinin2009b, mantz2010,
  rozo2010}.  Gas fraction measurements of baryonic to total mass have
also placed useful constraints on $\Omega_M$ when combined with D/H
abundance measurements of Ly$\alpha$ clouds and big bang
nucleosynthesis predictions \citep[e.g.,][]{white1993, david1995,
  white1995, neumann1997, squires1997, evrard1997, myers1997,
  ettori1999, mohr1999a, grego2001, laroque2006}. Such measurements
provided one of the strongest arguments that non-relativistic matter
alone does not close the universe at that time
\citep[e.g.,][]{viana1996, eke1996, kitayama1996, kitayama1997}.
Assuming the gas fraction does not evolve enables another method of
using gas fractions to place constraints on cosmological parameters
\citep[e.g.,][]{sasaki1996, pen1997, ettori2003, allen2004,
  allen2008}.  The determination of galaxy cluster temperatures is
particularly important because it is widely used to infer the
gravitational mass of clusters.

In the early 1970s, \citet{sunyaev1970, sunyaev1972} discussed the
inverse Compton scattering of cosmic microwave background (CMB)
photons off of energetic electrons of the hot cluster gas causing a
small ($\lesssim 1$mK) distortion in the CMB spectrum, now known as
the Sunyaev-Zel'dovich effect (SZE) \citep[for reviews
  see,][]{carlstrom2002, birkinshaw1999, rephaeli1995, sunyaev1980}.
The SZE is independent of redshift making it an attractive tool to
explore the high redshift universe.  Analysis of both SZE and X-ray
data of galaxy clusters provides a method of determining a direct
distance to galaxy clusters \citep{cavaliere1977, gunn1978, silk1978,
  cavaliere1978, birkinshaw1979}.  These distances are independent of
the extragalactic distance ladder and do not rely on standard candles
or rulers.  SZE/X-ray distances depend only on the properties of
highly ionized plasmas.  The promise of direct distances, in large
part, provided incentive to measure the small SZE signal.

Recently, anisotropies of the CMB, the Hubble diagram of type Ia
supernovae, and baryon acoustic oscillations in the galaxy power
spectrum have proved to be precise indicators of cosmology.  However,
galaxy cluster surveys probe the growth of structure, one of the few
methods to do so, and interpretation of survey yields has the
potential to precisely constrain the equation of state of the dark
energy \citep[e.g.,][]{bartlett1994, holder2000, haiman2001,
  majumdar2004}.  Current and future surveys of galaxy clusters in
radio \citep[ACT, SPT, APEX-SZ; e.g.,][]{fowler2010, hincks2009,
  fowler2007, carlstrom2010, lueker2010, reichardt2009}, X-ray
\citep[eROSITA; e.g.,][]{predehl2007, predehl2006}, and weak-lensing
\citep[Pan-STARRS, HSC, LSST, etc.; e.g.,][]{kaiser2004, miyazaki2006,
  ivezic2008} promise to provide unprecedented statistical samples of
galaxy clusters that will enable precision cosmology with galaxy
clusters.  Therefore it is important to consider possible systematics
inherent to using clusters as tools of cosmology.

Traditionally SZE/X-ray estimates of the Hubble parameter tend to be
low \citep[for details see, e.g.,][]{birkinshaw1999, carlstrom2002},
with a few exceptions \citep[e.g.,][]{mason2001}.  However, a recent
study of 38 galaxy clusters using data from \chandra\ and the
OVRO/BIMA SZE imaging project find $H_0 \sim 74$ km s$^{-1}$
Mpc$^{-1}$ \citep{bonamente2006}, consistent with other probes, such
as the {\it Hubble Space Telescope (HST)} $H_0$ key project \citep{freedman2001},
supernova results \citep[e.g.,][]{hicken2009}, and recent CMB primary
anisotropy results \citep[e.g.,][]{komatsu2009, komatsu2010}.
Previous work using the same SZE data but \rosat\ and \asca\ X-ray
data for 18 galaxy clusters find $H_0 \sim 60$ km s$^{-1}$ Mpc$^{-1}$
\citep[][hereafter R02]{reese2002}.

We explore the effects of the \chandra\ calibration on cluster X-ray
spectral properties and their subsequent effects on the inferred
$H_0$, through an X-ray spectral analysis of \chandra\ archival data
of SZE clusters.  We perform a uniform, systematic spectroscopic X-ray
analysis of 38 galaxy clusters, comparing the effects of three
different \chandra\ calibrations on galaxy cluster properties.  The
clusters are chosen from a sample used for SZE/X-ray distance
measurements \citep[][hereafter B06]{bonamente2006}, facilitating the
exploration of how the effects of the calibration affect the inferred
Hubble parameter.

B06 find much less model dependence than one might naively expect for
determinations of $H_0$ from SZE/X-ray determined distances and
therefore we focus on the simplest case, the isothermal $\beta$-model
\citep{cavaliere1976, cavaliere1978}.  In addition, a simple
isothermal model will make it easier to isolate the effects of the
different \chandra\ calibrations on individual spectral properties,
such as cluster gas temperature, $T_e$, and the X-ray cooling
function, $\Lambda$, and how these differences propagate into
cosmological parameter determinations.  Because $H_0 \propto T_e^2
/ \Lambda(T_e)$, the Hubble constant provides a good test case for
tracking the effects of cluster temperature on cosmological
parameters.

The rest of the paper is organized as follows. We begin with a short
pedagogical discussion on modeling of the X-ray emission from galaxy
clusters in Section~\ref{sec:model} with particular attention to its
potential systematic uncertainties.  \chandra\ data reduction and
spectroscopic analysis for the three \chandra\ calibrations are
discussed in Section~\ref{sec:chandra-tspec}.  Implications for the
Hubble constant are explored in Section~\ref{sec:H0} and more general
implications are discussed in Section~\ref{sec:discussion}.
\chandra\ calibrations 3.1.0, 4.1.4, and 4.2.2 are considered and are
referred to more simply as 3.1, 4.1, and 4.2.  Throughout this paper,
all uncertainties are at 68\% confidence and we adopt a flat,
$\Lambda$-dominated cosmology with $\Omega_m = 0.27$ and
$\Omega_\Lambda = 0.73$ consistent with recent {\it Wilkinson
  Microwave Anisotropy Probe (WMAP)} results \citep{komatsu2010,
  komatsu2009}.  As is the convention the Hubble constant is sometimes
expressed as $h$, defined as $H_0 = 100\ h$ km s$^{-1}$ Mpc$^{-1}$.

\section{Modeling X-ray surface brightness of galaxy clusters for 
estimating the Hubble constant} \label{sec:model}

\subsection{Theoretical modeling}
\label{subsec:theory_model}

We start with a pedagogical discussion of X-ray emission from galaxy
clusters in order to elucidate important details relevant to this work
and differences within the literature.  We denote the X-ray emissivity
[erg s$^{-1}$ cm$^{-3}$ keV$^{-1}$] at the source as
\begin{eqnarray}
\label{eq:xemissivity}
\frac{d^2 L_X}{dE_{\rm source} dV_{\rm source}}
\equiv \lambda (E_{\rm source},T_{\rm source}, Z_{\rm source}) n_e^2,
\end{eqnarray}
where $L_X$ is X-ray luminosity, $E$ is photon energy, $V$ is volume,
$T=T_e$ is the cluster temperature, $Z$ is metallicity relative to
solar, $n_e$ is the electron density, and we use the index ``source''
for clarity to indicate variables defined at the source position (with
redshift $z$). The function $\lambda (E_{\rm source}; T_{\rm source},
Z_{\rm source})$ [erg s$^{-1}$ cm$^{3}$ keV$^{-1}$] depends on the
temperature $T_{\rm source}$ and the metal abundance $Z_{\rm source}$
of the source, and is given by the theoretical model.

Consider X-ray photons isotropically emitted from the source of physical
size $\Delta x \Delta y \Delta \ell$ (with $\ell$ chosen to be along
the line of sight of the observer at $z=0$) with energy range $E_{\rm
source} \sim E_{\rm source}+\Delta E_{\rm source}$ for a time interval
$\Delta t_{\rm source}$. Then the total number of the photons is given
by
\begin{eqnarray}
\label{eq:photon-at-source}
{\cal N}_X^{{\rm source}} =
\frac{1}{E_{\rm source}}
\frac{d^2 L_X}{dE_{\rm source} dV_{\rm source}}
\times \Delta E_{\rm source} \Delta t_{\rm source} \Delta x
\Delta y \Delta \ell .
\end{eqnarray}
 
The ideal observer (with perfect efficiency) receives the same number of
photons from the solid angle of the source $\Delta \Omega_{\rm obs}
\equiv \Delta \theta_x \Delta \theta_y = \Delta x \Delta y /d_{\rm A}^2$
for a time interval $\Delta t_{\rm obs}$ with energy range $E_{\rm obs}
\sim E_{\rm obs} + \Delta E_{\rm obs}$, where those variables with index
``obs'' are defined at the observer's frame, and $d_{\rm A}$ is the angular
diameter distance to the source. If we denote the ``differential''
surface brightness of the source as
\begin{eqnarray}
\frac{d S_X^{\rm theory}}{dE_{\rm obs}}
\qquad
[ {\rm erg} ~  {\rm s}^{-1} ~ {\rm cm}^{-2} ~
{\rm arcmin}^{-2} ~ {\rm keV}^{-1}] ,
\end{eqnarray}
then the number of photons in the observer's frame is given by
\begin{eqnarray}
\label{eq:photon-at-observer}
{\cal N}_X^{{\rm obs}} =
4\pi d_{\rm com}^2 
\frac{1}{E_{\rm obs}}
\frac{d S_X^{\rm theory}}{dE_{\rm obs}}
\times \Delta E_{\rm obs} \Delta t_{\rm obs} \Delta \Omega_{\rm obs},
\end{eqnarray}
where $d_{\rm com}$ is the comoving distance to the source at redshift
$z$.  Since the number of photons in both frames is invariant, ${\cal
  N}_X^{\rm obs} = {\cal N}_X^{\rm source}$, Equations
(\ref{eq:photon-at-source}) and (\ref{eq:photon-at-observer})
are equal and we obtain
\begin{eqnarray}
\frac{d S_X^{\rm theory}}{dE_{\rm obs}}
&=& \frac{1}{4\pi  d_{\rm com}^2 }
\frac{E_{\rm obs}}{E_{\rm source}}
\frac{d^2 L_X}{dE_{\rm source} dV_{\rm source}}
\frac{\Delta t_{\rm source}}{\Delta t_{\rm obs}}
\frac{\Delta E_{\rm source}}{\Delta E_{\rm obs}}
\frac{\Delta x \Delta y}{\Delta \Omega_{\rm obs}} \Delta \ell \cr
&=& \frac{1}{4\pi  d_{\rm com}^2 }
\frac{1}{1+z}
\frac{d^2 L_X}{dE_{\rm source} dV_{\rm source}}
\frac{1}{1+z} 
(1+z)
\, 
d_{\rm A}^2 \Delta \ell \cr
&=& \frac{1}{4\pi (1+z)} \left(\frac{d_{\rm A}}{d_{\rm com}}\right)^2  
\frac{d^2 L_X}{dE_{\rm source} dV_{\rm source}} \Delta \ell \cr
&=& \frac{1}{4\pi (1+z)^3} 
\frac{d^2 L_X}{dE_{\rm source} dV_{\rm source}} \Delta \ell,
\end{eqnarray}
where we use $E_{\rm source}=(1+z)E_{\rm obs}$, 
$\Delta t_{\rm source}= \Delta t_{\rm obs}/(1+z)$, and
$d_{\rm A}= d_{\rm com}/(1+z)$.

If the profile along the line of sight is properly taken into account,
$\Delta \ell$ in the above equation will be replaced by the
integration along the line of sight:
\begin{eqnarray}
\label{eq:dsdeobs}
\frac{d S_X^{\rm theory}}{dE_{\rm obs}}
&=& \frac{1}{4\pi (1+z)^3} 
\int d\ell \frac{d^2 L_X}{dE_{\rm source} dV_{\rm source}} \cr
&=& \frac{1}{4\pi (1+z)^3} 
\int d\ell \, 
\lambda (E_{\rm source},T_{\rm source}, Z_{\rm source}) n_e^2 .
\end{eqnarray}
Then the surface brightness of the source, $S_X^{\rm theory}$ [erg
s$^{-1}$ cm$^{-2}$ arcmin$^{-2}$], defined at the observer's frame is
given by
\begin{eqnarray}
\label{eq:sx_theory}
S_X^{\rm theory} 
&=& \int dE_{\rm obs} \frac{d S_X^{\rm theory}}{dE_{\rm obs}}
= \frac{1}{4\pi (1+z)^3} 
\int dE_{\rm obs} \int d\ell 
\frac{d^2 L_X}{dE_{\rm source} dV_{\rm source}} \cr
&=& \frac{1}{4\pi (1+z)^4} 
\int d\ell 
\int dE_{\rm source} \, 
\lambda (E_{\rm source},T_{\rm source}, Z_{\rm source}) n_e^2 
\cr
&=& \frac{1}{4\pi (1+z)^4} 
\int d\ell \, 
\Lambda^{\rm theory} (T_{\rm source}) n_e^2 ,
\end{eqnarray}
where the cooling function $\Lambda^{\rm theory}(T_{\rm source})$ [erg
s$^{-1}$ cm$^{3}$] is given by
\begin{eqnarray}
\label{eq:lambda-theory}
\Lambda^{\rm theory} (T_{\rm source}) 
\equiv \int dE_{\rm source} \, 
\lambda (E_{\rm source},T_{\rm source}, Z_{\rm source}).
\end{eqnarray}
Note that $\Lambda^{\rm theory}(T_{\rm source})$ indeed depends on
$Z_{\rm source}$ as well, but we do not write it explicitly for
simplicity of notation.

\subsection{From observed photon counts to surface brightness}
\label{subsec:det_to_cgs}

In real observations, we have to take account of the overall response
function of the detector and the telescope, whose effect is expressed
by the energy-dependent effective area $A(E_{\rm obs})$ [cm$^2$]. Then
the observed number of photons per unit time per unit solid angle
counted on the detector, $d^2{\cal N}_X^{\rm obs}/(dt_{\rm
  obs}d\Omega_{\rm obs})$ [counts s$^{-1}$ arcmin$^{-2}$], is modeled
as
\begin{eqnarray}
\label{eq:nxobs}
\frac{d^2{\cal N}_X^{{\rm obs}}}{dt_{\rm obs}d\Omega_{\rm obs}}
&=& \int dE_{\rm obs} \, \frac{1}{E_{\rm obs}}
\frac{d S_X^{\rm theory}}{dE_{\rm obs}} A(E_{\rm obs}) \cr
&=& \int dE_{\rm source} \, \frac{1}{E_{\rm source}}
\frac{d S_X^{\rm theory}}{dE_{\rm obs}} A(E_{\rm obs}) \cr
&=& \frac{1}{4\pi (1+z)^3}\int n_e^2  d\ell \cr
&& \qquad\times \int dE_{\rm source} \, 
\frac{\lambda (E_{\rm source},T_{\rm source}, Z_{\rm source})}
{E_{\rm source}}
A\left(\frac{E_{\rm source}}{1+z}\right).
\end{eqnarray}
Spectral analysis of the observed X-ray photons yields $T_{\rm
  source}$ and $Z_{\rm source}$ that best-fit the data.

Let us define the {\it effective} cooling function as
\begin{eqnarray}
\label{eq:Lambdaeff}
&& \Lambda^{\rm eff}(T_{\rm source}) 
\equiv \frac{1}{A(E_{\rm obs}=E_{\rm fid})} \cr
&\times& \int_{E_{1,\rm obs}(1+z)}^{E_{2,\rm obs}(1+z)} dE_{\rm source} \, 
\frac{\lambda (E_{\rm source},T_{\rm source}, Z_{\rm source})}
{E_{\rm source}}
A\left(\frac{E_{\rm source}}{1+z}\right) ,
\end{eqnarray}
where $E_{\rm fid}$ is some fiducial energy, $E_{\rm fid} = 1$ keV for
the \chandra\ analyses presented and compared in this work, and
$E_{1,\rm obs}$ and $E_{2,\rm obs}$ correspond to the observed energy
band of the detector (or of the analysis).

Before proceeding further, a few comments should be added here. First,
this quantity has units of [erg s$^{-1}$ cm$^{3}$ keV$^{-1}$] =
[counts s$^{-1}$ cm$^{3}$], different from that of $\Lambda^{\rm
  theory}$ [erg s$^{-1}$ cm$^{3}$]. In addition to covering different
energy ranges, \rosat\ and \chandra\ treat exposure maps differently,
which leads to different treatments of the ``cooling function''.
Therefore one must be careful when comparing emissivity and surface
brightness results even in cgs units.  Of course, a ``count'' in
different observatories corresponds to different energy photons, on
average, so ``detector'' unit results most often can not be compared
directly either.  Choosing $E_{\rm fid} = 1$ keV in
B06 \nocite{bonamente2006} is effectively arbitrary but does roughly
correspond to where \chandra\ is most sensitive and corresponds to the
energy at which the exposure maps used in their analysis were
computed.  We follow suit and adopt $E_{\rm fid} = 1$ keV for a more
straight-forward comparison.

Finally the {\it effective} cooling function depends on $Z_{\rm
  source}$ and $z$ in addition to $T_{\rm source}$, but we write it as
$\Lambda^{\rm eff}(T_{\rm source}) $ for simplicity.  Equations
(\ref{eq:nxobs}) and (\ref{eq:Lambdaeff}) are now combined to give
\begin{eqnarray}
\label{eq:nxobsperarea}
\frac{d^2{\cal N}_X^{{\rm obs}}}{dt_{\rm obs}d\Omega_{\rm
    obs}}\frac{1}{A(E_{\rm obs}=E_{\rm fid})}= \frac{1}{4\pi
  (1+z)^3} \int n_e^2 d\ell \times \Lambda^{\rm eff}(T_{\rm source}) .
\end{eqnarray}
Comparing this equation with Equation (\ref{eq:sx_theory}), we find
that if we define the quantity:
\begin{eqnarray}
\label{eq:Sx_obs-def}
{\cal S}_X^{\rm obs}
\equiv 
\frac{d^2{\cal N}_X^{{\rm obs}}}{dt_{\rm obs}d\Omega_{\rm
    obs}}\frac{1}{A(E_{\rm obs}=E_{\rm fid})},
\end{eqnarray}
then we obtain
\begin{eqnarray}
\label{eq:Sx_obs}
{\cal S}_X^{\rm obs}
= \frac{1}{4\pi (1+z)^3} \int n_e^2  d\ell
\times \Lambda^{\rm eff}(T_{\rm source}) .
\end{eqnarray}
Again it should be noted that ${\cal S}_X^{\rm obs}$ [erg s$^{-1}$ cm$^{-2}$
  arcmin$^{-2}$ keV$^{-1}$] (= [counts s$^{-1}$ cm$^{-2}$
  arcmin$^{-2}$]) is not directly related to the surface brightness
$S_X^{\rm theory}$, but this notation follows that used in the
literature.\footnote{In R02, \nocite{reese2002} some effort was made to
  distinguish between the observed, denoted as ``detector'', and
  theoretical, denoted as ``cgs'', quantities and both were presented
  while the B06 \nocite{bonamente2006} analysis presents only observed
  (detector) values.}

In any case, if the effective cooling function is independent of the
line-of-sight integral of Equation (\ref{eq:Sx_obs}), such as in the
isothermal case considered here, we can estimate the density squared
integrated along the line of sight as
\begin{eqnarray}
\frac{1}{4\pi (1+z)^4} \int n_e^2  d\ell
= \frac{{\cal S}_X^{\rm obs} ~[{\rm counts ~ s}^{-1}~ {\rm cm}^{-2}
~ {\rm arcmin}^{-2}]}
{(1+z) \Lambda^{\rm eff}(T_{\rm source}) 
~[{\rm counts ~ s}^{-1}~ {\rm cm}^3]} .
\end{eqnarray}
Even though ${\cal S}_X^{\rm obs}$ and $\Lambda^{\rm eff}(T_{\rm source})$ are
not the direct observational counterparts of $S_X^{\rm theory}$ and
$\Lambda^{\rm theory}(T_{\rm source})$, respectively, their ratio may be
used to estimate the left-hand-side in the above equation.

\subsection{Breakdown of factors important for determining $H_0$ with
  different calibrations}
\label{subsec:h0_breakdown}

As discussed previously, the X-ray emission is
\begin{eqnarray}
{\cal S}_X^{\rm obs} \propto \int n_e^2 \, \Lambda^{\rm eff} \, d\ell 
\propto n_e^2 \, \Lambda^{\rm eff} \, \Delta \ell.
\end{eqnarray}
Since the observed SZE temperature decrement is proportional to the
pressure integrated along the line-of-sight,
\begin{eqnarray}
\Delta T_{\rm SZE}
\propto  f_{(\nu, T_{\rm source})}
\int n_e T_{\rm source} \, d \ell 
\sim  f_{(\nu, T_{\rm source})} n_e T_{\rm source} \, \Delta \ell ,
\end{eqnarray}
the angular diameter distance, $d_A$, can be estimated by taking
advantage of the different dependencies on $n_e$.  The frequency
dependence of the SZE, $f_{(\nu, T_e)}$, also depends on temperature
when relativistic effects are considered \citep[e.g.,][]{itoh1998,
  challinor1998}. Eliminating $n_e$ in favor of $\Delta \ell$ yields
\begin{eqnarray}
d_A \propto \Delta \ell \propto \left ( \frac{\Delta T_{\rm
    SZE}}{T_{\rm source}\, f_{(\nu, T_{\rm source})}} 
\right )^2 \frac{\Lambda^{\rm eff}}{{\cal S}_X^{\rm obs}}.
\end{eqnarray}
Therefore the estimated Hubble constant should be proportional to
\begin{eqnarray}
H_{0, \rm est} \propto \Delta \ell^{-1} \propto \frac{T_{\rm source}^2
  \, {\cal S}_X^{\rm obs} \, f^2_{(\nu, T_{\rm source})}}{\Lambda^{\rm
    eff}(T_{\rm source})} \propto \frac{T_{\rm source}^2 f^2_{(\nu,
    T_{\rm source})}}{\Lambda^{\rm eff}(T_{\rm source}) \, A(E_{\rm
    obs}=E_{\rm fid})} ,
\end{eqnarray}
where we assume that the values of $\Delta T_{\rm SZE}$ and ${\cal
  N}_X^{\rm obs}$ are not affected by the change of the calibration.
Details on the calculation of distances from the analysis of X-ray and
SZE data can be found elsewhere \citep[e.g.,][]{birkinshaw1991,
  birkinshaw1999, reese2002}.  The dependence of $f_{(\nu, T_{\rm
    source})}$ on $T_e$ is weak and we neglect it in the rest of this
section, although we do include it in the distance and Hubble constant
calculations from the data (Section~\ref{sec:H0}).

Thus the ratio of the estimates of $H_0$ from two different
calibrations, 1 and 2, is finally written as
\begin{eqnarray}
\label{eq:H0-new-old}
\frac{H_{0, 2}}{H_{0, 1}}
&=& \frac{T_{2}^2}{\Lambda_{2}^{\rm eff}(T_{2}) 
\, A_{2}(E_{\rm fid})} 
\frac{\Lambda_{1}^{\rm eff}(T_{1}) 
\, A_{1}(E_{\rm fid})} 
{T_{1}^2} \cr
&=& \left ( \frac{T_{2}}{T_{1}} \right )^2
\frac{\Lambda_{1}^{\rm eff}(T_{1}) }
{\Lambda_{2}^{\rm eff}(T_{2})}
\frac{A_{1}(E_{\rm fid})} {A_{2}(E_{\rm fid})}.
\end{eqnarray}
Equation (\ref{eq:H0-new-old}) can be
rewritten in a form useful in understanding
the error budget of the estimate of $H_0$
\begin{eqnarray}
\label{eq:H0-new-old-2}
\frac{H_{0, 2}}{H_{0, 1}}
= \left ( \frac{T_{2}}{T_{1}} \right )^2
\frac{\Lambda_{2}^{\rm eff}(T_{1}) }
{\Lambda_{2}^{\rm eff}(T_{2})}
\frac{\Lambda_{1}^{\rm eff}(T_{1}) }
{\Lambda_{2}^{\rm eff}(T_{1})}
\frac{A_{1}(E_{\rm fid})} {A_{2}(E_{\rm fid})} ,
\end{eqnarray}
where $\Lambda_{2}^{\rm eff}(T_{1})$ is the effective cooling function
using the temperature from observation 1 but the effective area
corresponding to observation 2.  Because it enters the
$H_0$ calculation squared, the effects of the calibration on $T_e$
will have the greatest impact.  For instance, a $\sim 10$\% change in
$T_e$ would result in a $\sim 20$\% change in the derived Hubble
parameter.

The product of the third and fourth factors indicates the effect of the
different effective areas evaluated at the same temperature (and same
abundance). More specifically it is given by
\begin{eqnarray}
&& \frac{\Lambda_{1}^{\rm eff}(T_{1}) }
{\Lambda_{2}^{\rm eff}(T_{1})}
\frac{A_{1}(E_{\rm fid})} {A_{2}(E_{\rm fid})} \cr
&=&
\frac
{\displaystyle 
\int_{E_{1,\rm obs}(1+z)}^{E_{2,\rm obs}(1+z)} dE_{\rm source} \, 
\frac{\lambda (E_{\rm source},T_{1}, Z_{1})}
{E_{\rm source}}
A_{1}\left(\frac{E_{\rm source}}{1+z}\right) }
{\displaystyle 
\int_{E_{1,\rm obs}(1+z)}^{E_{2,\rm obs}(1+z)} dE_{\rm source} \, 
\frac{\lambda (E_{\rm source},T_{1}, Z_{1})}
{E_{\rm source}}
A_{2}\left(\frac{E_{\rm source}}{1+z}\right) }.
\end{eqnarray}
The above factors may change the inferred Hubble parameter by several
percent (see Figure~\ref{fig:ea} in
Section~\ref{subsec:spec_analysis}).

\section{Spectroscopic temperature of SZE clusters with \chandra}
\label{sec:chandra-tspec}

We determine the spectroscopic temperature of all 38 clusters used in
the B06 \nocite{bonamente2006} analysis with three different versions of
the calibration database and corresponding ciao versions: calibration
versions 3.1.0, 4.1.4, and 4.2.2 with ciao versions 3.4, 4.1.2, and
4.2.  These calibrations will be referred to as 3.1, 4.1, and 4.2
hereafter.

The most significant change between the 3.1 and 4.1 calibrations is
the updated high-resolution mirror assembly (HRMA) effective area in
version 4.1.  A new effective area model was developed after
revisiting the ground calibration data and ray-trace model, prompted,
in large part, by differences in inferred temperatures of massive
galaxy clusters between \chandra\ and \xmm.  The newer effective area
at low energy ($E<5$ keV) is $\lesssim 10$\% lower than that of the
older
calibration.\footnote[1]{http://cxc.harvard.edu/ciao/releasenotes/ciao\_4.1\_release.html}
The main change in calibration version 4.2 is the AXAF CCD imaging
spectrometer (ACIS) contamination
model.\footnote[2]{http://cxc.harvard.edu/ciao/releasenotes/ciao\_4.2\_release.html}
Based on external calibration source (ECS) measurements, separate
ACIS-I and ACIS-S models were developed.  An offset was added to
normal exponential evolution of the contamination model for
ACIS-S,\footnote[3]{http://cxc.harvard.edu/cal/memos/contam\_memo.pdf}
while there is no such offset applied for ACIS-I.  We note that there
are a lot of other minor changes between 4.1 and earlier versions, and
between 4.1 and 4.2. These more minor changes also have the potential
to affect changes in the estimated cluster temperatures but to a much
lesser extent than the changes outlined above.

\subsection{Data Reduction}
\label{subsec:obs_data_reduce}

All available archival \chandra\ data for the 38 galaxy clusters in
our sample are used in the analysis.  This includes all data used in
B06 \nocite{bonamente2006} as well as observations available at the
time but not included and new observations since that time that are in
the public archive.  There are 87 observations among the 38 galaxy
clusters, not including two observations of A0267 that have $\lesssim
1$ ks of observation time that were not used.  The data are summarized
in Table~\ref{tab:data}.  Cluster position, redshift, \ion{H}{1}
column density \citep{dickey1990}, observation identification number,
ACIS configuration (ACIS-I or ACIS-S), and livetime (effective
integration time) are shown.  Redshift references may be found in
B06.\nocite{bonamente2006}

The data are reduced with each of the three different versions
of the calibration database 3.4, 4.1, and 4.2.  The data are
processed starting with the level 1 events data, removing the cosmic
ray afterglow correction, and generating a new bad pixel file that
accounts for hot pixels and cosmic ray afterglows.  Using the newly
generated bad pixel file, the charge transfer inefficiency correction,
time-dependent gain adjustment, and other standard corrections are
applied to the data.  The data are filtered for \asca\ grades 0, 2, 3,
4, 6 and status=0 events and the good time interval data provided with
the observations are applied.  Periods of high background count rate
are excised using an iterative procedure involving creating light
curves in background regions with 259 s bins (following the ACIS
``Blank-Sky'' Background File reduction), and excising time intervals
that are in excess of 3 $\sigma$ (=rms) from the mean background count
rate.  This sigma clipping procedure is iterated until all remaining
data lie within 3 $\sigma$ of the mean.  The final events list is
limited to energies 0.7-7.0 keV to exclude the low- and high-energy
data that are more strongly affected by calibration uncertainties.

Lightcurve filtering is performed on background regions chosen for
each observation.  Three circular regions, masking out any intervening
point sources, are used as the background regions.  Investigation of
deep blank-sky exposures shows that front-illuminated (FI) and
back-illuminated (BI) chips have different responses
\citep{bonamente2004}.  In particular, the BI chips are basically
constant over the chip but there is a gradient in the FI chips that
depends on the distance from the readout nodes.  For observations
taken in the ACIS-S configuration, the cluster observation falls on a
BI chip.  In these cases, background regions are chosen at the
periphery of the BI chip, away from the cluster emission.  For
observations taken in the ACIS-I configuration, there are four main FI
chips.  The cluster falls on one of those chips (I3).  Background
regions are chosen from the three remaining chips (I0, I1, and I2),
one region for each chip, using locations at the same distance from
the readout nodes as the cluster.

To facilitate point source detection, images and exposure maps are
constructed.  Images are created by binning the data by a factor of 4, resulting in a pixel size of 1\arcsec.97.  Exposure maps are
constructed for each observation at an energy of 1 keV.  For clusters
with multiple data sets, the observation with the longest exposure
time is used for point source detection.  A wavelet based source
detector is used to find and generate a list of potential point
sources.  The list is examined by eye, removing bogus or suspect
detections, and then used as the basis for our point source mask.

Spectra are extracted from the cluster using a region that encompasses
95\% of the galaxy cluster counts, accounting for point sources that
fall within the region of interest.  The same wavelet based source
detection algorithm as for point source detection is used to find the
cluster by having it search for large scales (compared to the
point-spread function (PSF)) and determine the initial extraction
region.  The center of the ellipse returned by the source detector is
adopted as the center of the extraction region and the semimajor and
semiminor axes of the ellipse are combined in quadrature and its
square root used as the radius for the circular extraction region.
Background subtracted counts are then computed in annuli out to the
initial extraction radius and the region encompassing 95\% of the
cluster counts is adopted as the final extraction radius.  This
provides a formulaic method of constructing spectral extraction
regions.

Calibration version 4.1 is used to construct background and
spectral extraction regions that are then used for the other two
reductions.  This provides a uniform reduction and systematic analysis
procedure that isolated the effects of the calibration.  Using
these regions, spectra are extracted and responses computed for each
observation.  Multiple observations for a given cluster each have
spectra and response files that are then simultaneously fit to a
thermal spectrum.

\subsection{Spectral Analysis}
\label{subsec:spec_analysis}

Following \citet{bonamente2006, bonamente2004}, XSPEC
\citep{arnaud1996, dorman2001} is used to model the intracluster
medium (ICM) with a Raymond$-$Smith spectrum \citep{raymond1977}
accounting for galactic extinction, with solar abundances of
\citet{feldman1992}, and cross sections of \citet{balucinska1992} with
an updated He cross section \citep{yan1998}.  The analysis uses data
in the 0.7-7.0 keV energy range.  We use the ``cstat'' statistic in
XSPEC, essentially the so-called Cash-statistic \citep{cash1979}, to
properly account for low count spectral bins.  Using the $\chi^2$
statistic in such cases will result in biased spectral parameters.  We
perform a Markov Chain Monte Carlo (MCMC) analysis, allowing us to
compute proper uncertainties on the X-ray cooling function,
$\Lambda_{\rm eff}$, which also enters the distance calculation and is
derived from the spectral parameters from the MCMC chain.

A simultaneous fit is performed to all data sets of a given cluster.
Responses for each individual cluster observation may be quite
different even for the same cluster.  In particular, the FI and BI
ACIS chips show considerably different responses.  Therefore, for
clusters with multiple observations, each individual observation is
modeled independently, using their individual response and background
spectra.  Normalizations of the model are allowed to vary for each
data set while the temperature and abundance are linked between the
different data sets.  Redshift and column density are held fixed,
adopting the \citet{dickey1990} $N_{\mathrm{H}}$ values (See
Table~\ref{tab:data}).

\begin{figure}[!tbh]
  \centerline{
    \includegraphics[width=85mm]{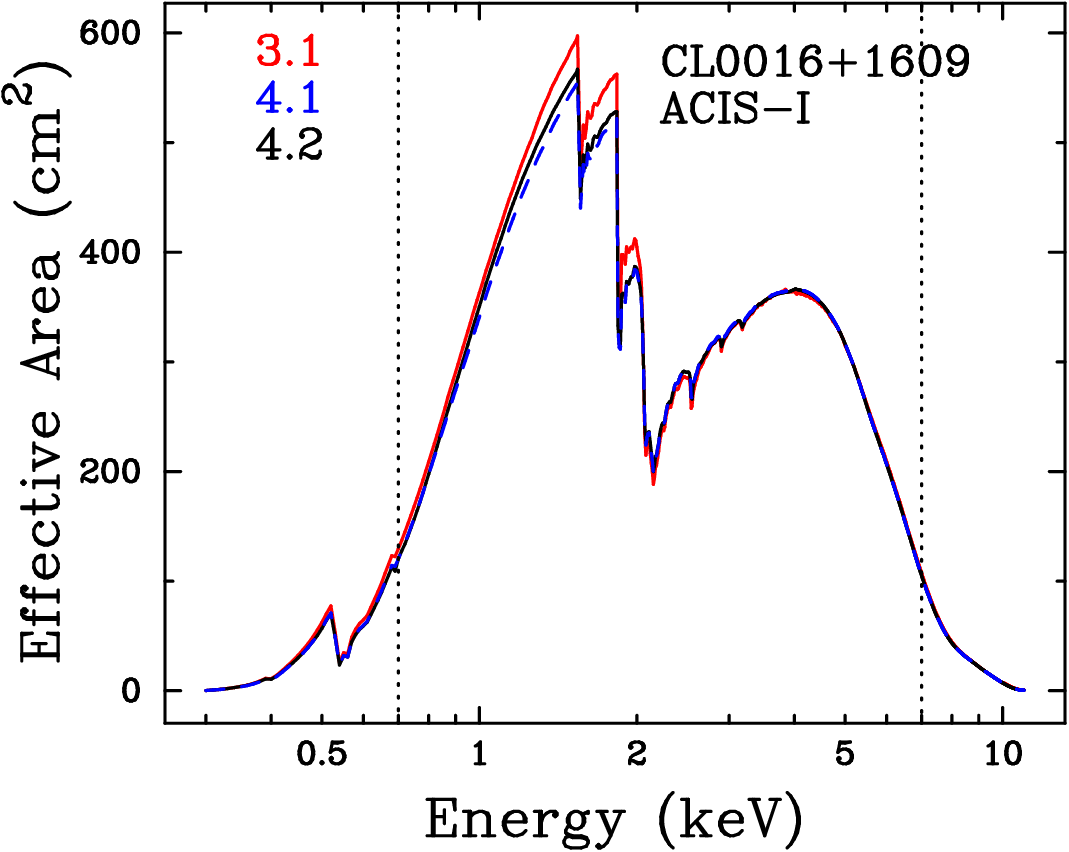}
    \includegraphics[width=85mm]{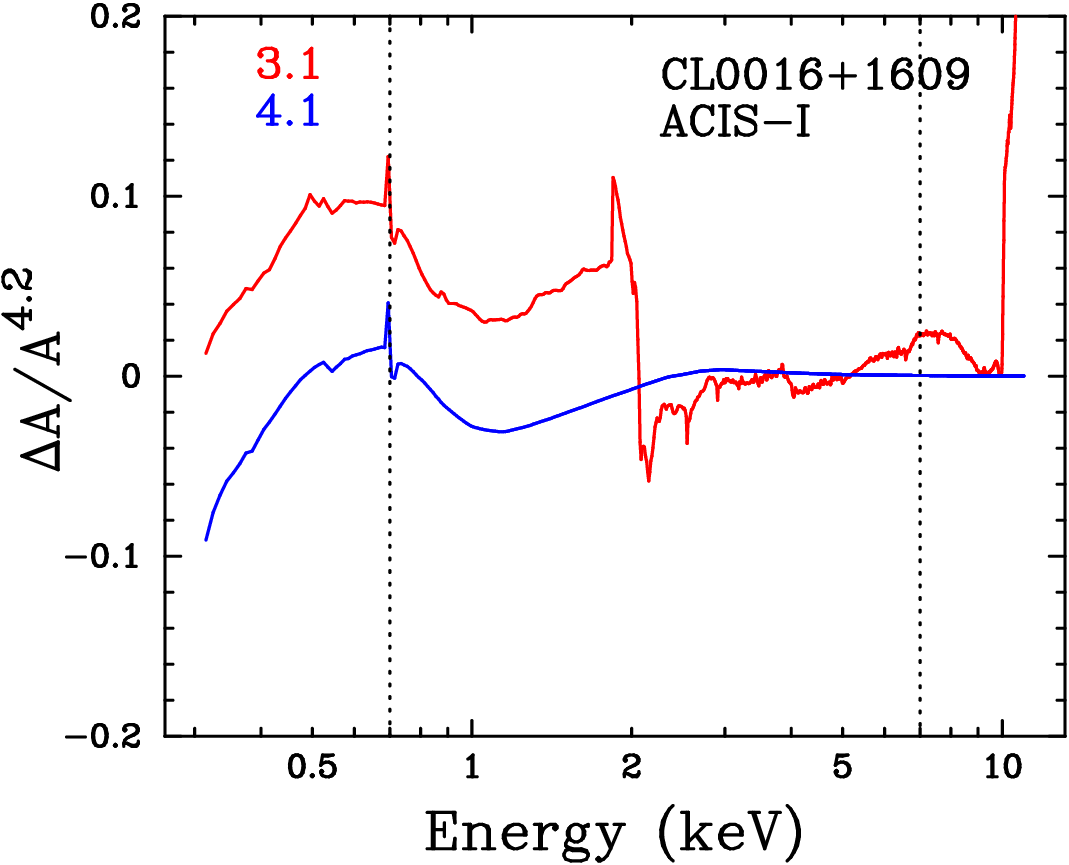}
  }
  \centerline{
    \includegraphics[width=85mm]{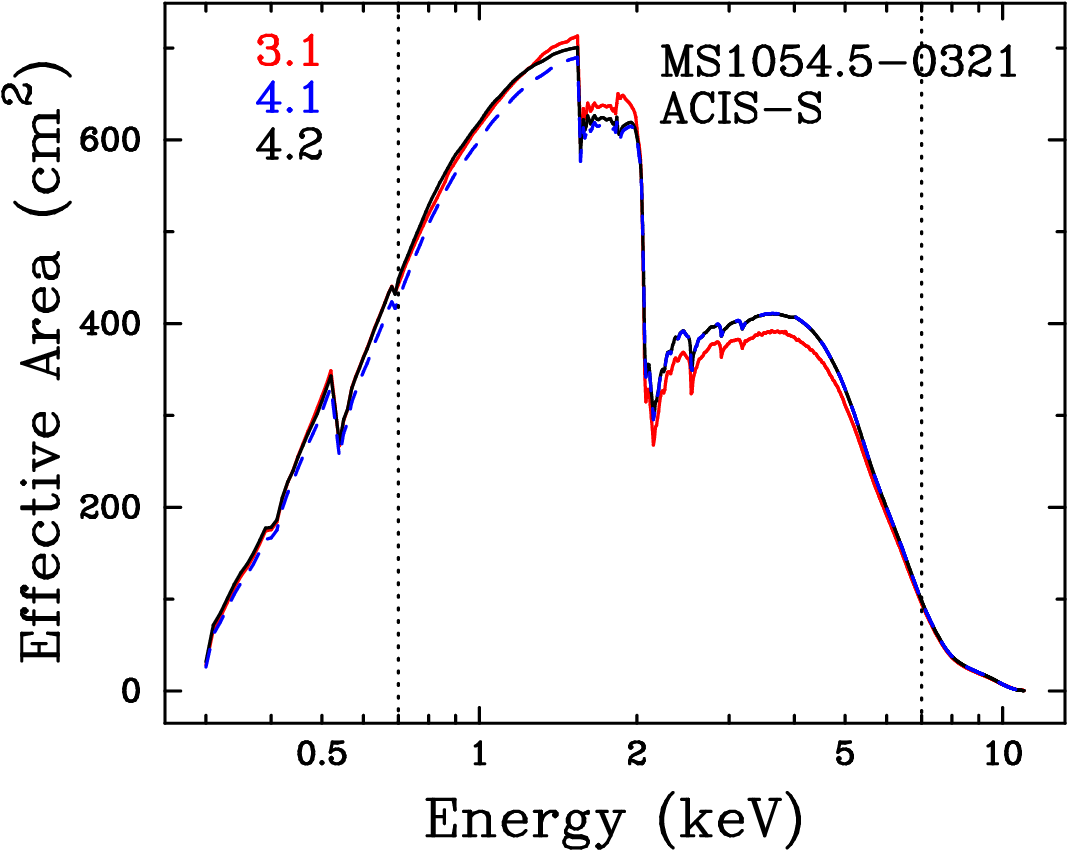}
    \includegraphics[width=85mm]{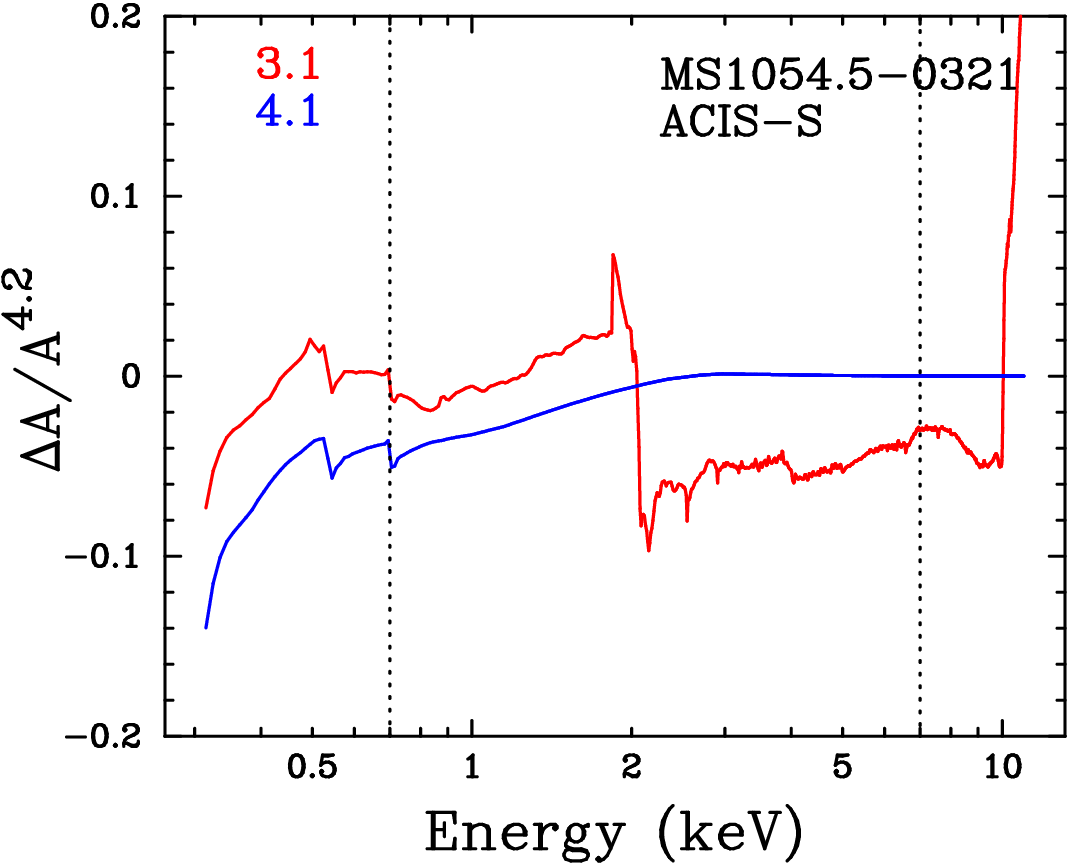}
  }
  \caption{Comparison of the effective areas for the 3.1 (red), 4.1
    (blue), and 4.2 (black) calibrations for examples of ACIS-I
    (CL0016+16; top) and ACIS-S (MS1054.5-0321; bottom) observations.
    The fractional residuals from the 4.2 effective area are shown in
    the right hand panels, $\Delta A/A^{4.2}=(A^{x}-A^{4.2})/A^{4.2}$,
    where $x$ is either 3.1 (red) or 4.1 (blue).  In addition to the
    changes seen between 1 and 2 keV in both ACIS-I and ACIS-S
    observations, ACIS-S observations also exhibit appreciable
    differences between the 3.1 and 4.1/4.2 calibrations in the 2$-$5
    keV range.  The vertical dotted lines mark the 0.7-7.0 keV energy
    range used in this analysis.
    \label{fig:ea}}
\end{figure}

Markov chains are run for 100,000 iterations.  We drop the initial
5000 iterations for the burn-in period but the results are insensitive
to that choice.  The X-ray cooling function, $\Lambda_{\rm eff}$, is
computed at each step in the Markov chain, enabling the folding of
uncertainties of the spectral parameters to the computed
$\Lambda_{\rm eff}$.  Best-fit parameters and confidence intervals are
computed from the cumulative distribution with 50\%, 16\%, and 84\%
probability, corresponding to the median and 68\% confidence interval.
The resultant probability distribution functions for each fit and
derived parameter of each cluster are visually inspected.
Convergence and mixing are checked with the Geweke $Z$-statistic
\citep{geweke1992}.

For clusters with multiple observations, a weighted average effective
area as a function of energy is used to compute the X-ray cooling
function.  The weighted average effective area, $A_{\mathrm{avg}}(E)$, is the
average effective area weighted by the livetime (effective integration
time) of each observation, $A_{\mathrm{avg}}(E)=(\sum A_i(E) * t_i) / \sum
t_i$, where $A_i(E)$ is the effective area of observation $i$, $t_i$
is the livetime of observation $i$, and the sum is over the number of
observations for that cluster.  The weighted average effective area is
used to compute $\Lambda_{\rm eff}$ at each step in the Markov chain and
determine the best-fit value and 68\% confidence interval.

\subsection{Spectral Results}
\label{sub:spec_results}

The newest calibration version, 4.2, will be used as the baseline for
comparison with the other two calibrations throughout the analysis.
We define the mean ratio of parameter $P$ between calibrations as
\begin{eqnarray}
\left < \frac{P^{x}}{P^{4.2}} \right > \equiv
\frac{1}{N_{\mathrm{cl}}} 
\sum_{i=1}^{N_{\mathrm{cl}}} \frac{P^{x}_i}{P^{4.2}_i}
\label{eq:mean_ratio_def}
\end{eqnarray}
where $i$ denotes the value for each individual cluster, $x$ refers to
3.1 or 4.1, and $N_{\mathrm{cl}}$ is the number of clusters used in the
calculation.

\begin{figure}[!tbh]
\centerline{
  \includegraphics[width=85mm]{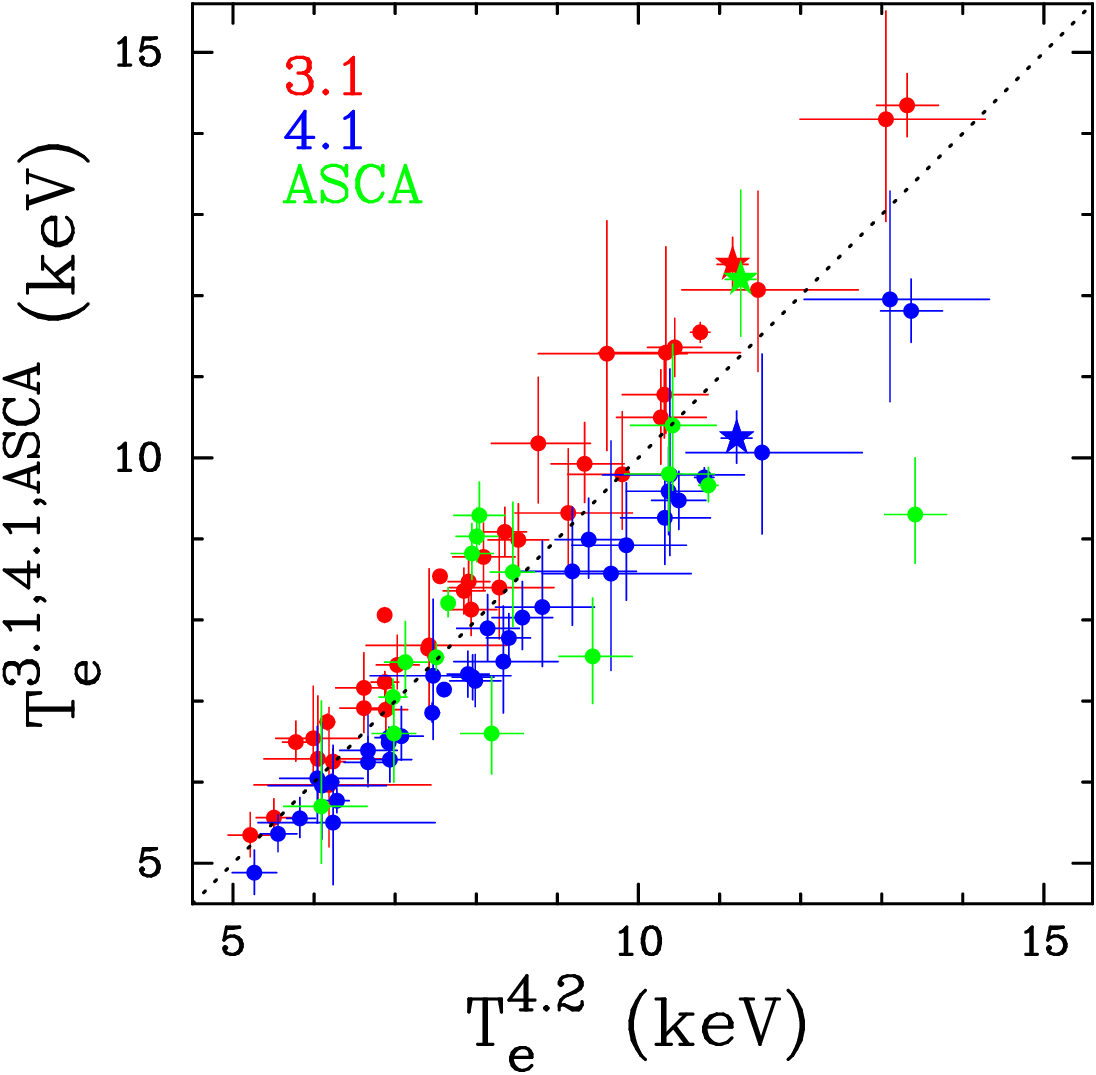}
  \includegraphics[width=85mm]{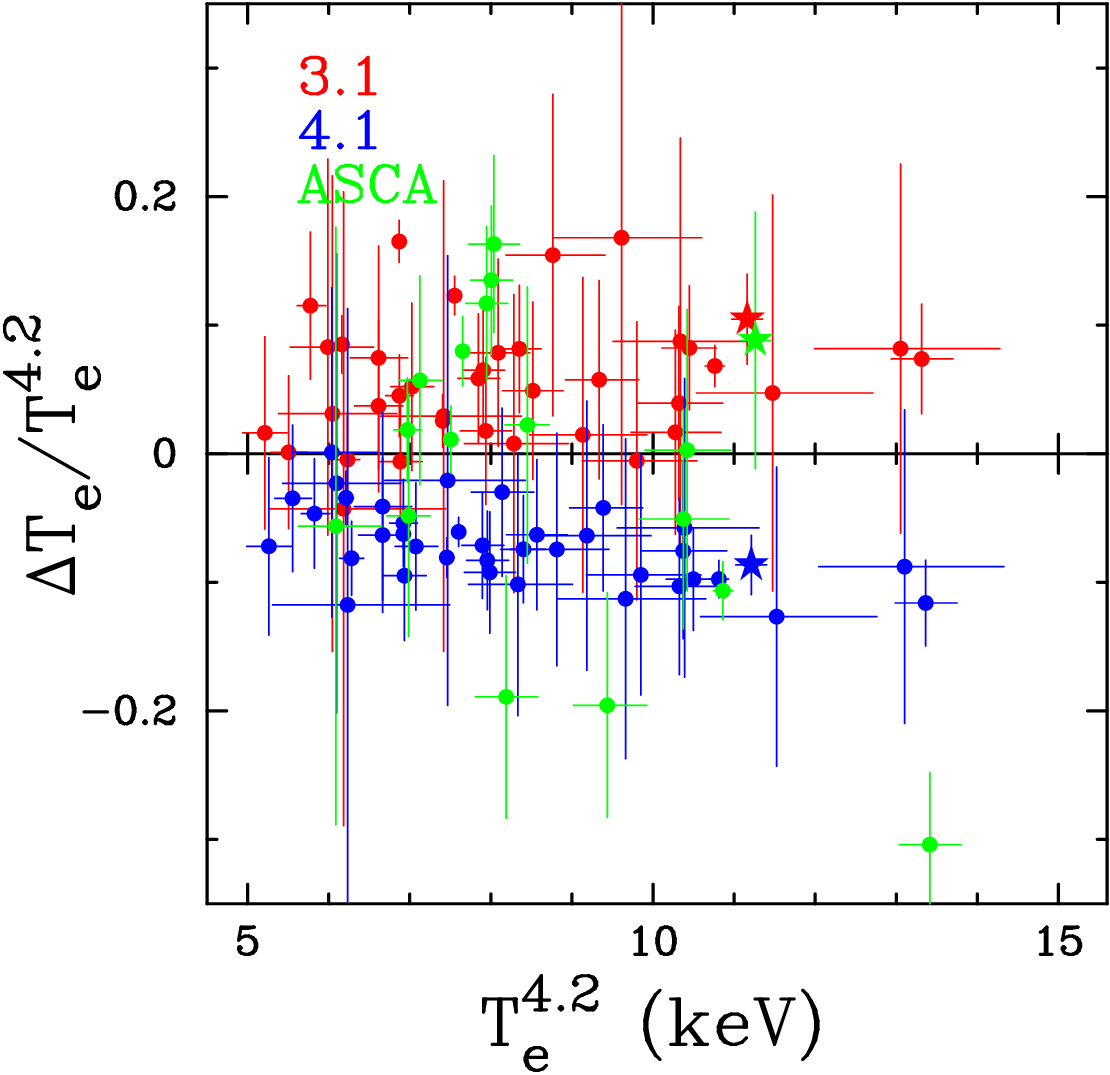}
}
\caption{Comparison of different temperature measurements (left) and
  the corresponding fractional residuals from the 4.2 calibration
  results (right).  Temperature determinations using the 3.1
  calibration (red) and 4.1 (blue) calibration are compared against
  temperatures using the newest 4.2 version.  Also plotted are the
  \asca\ (green) temperatures adopted in R02 
  for comparison of the 17 clusters that overlap with this work. The
  dotted black line shows the equality relation.  Residuals are
  defined as $\Delta T_e / T_e^{4.2} = (T_e^x-T_e^{4.2})/T_e^{4.2}$,
  where $x$ is the 3.1 (red), 4.1 (blue), or \asca\ (green) results.
  Error bars show 68\% confidence statistical uncertainties and the
  A2163 results are denoted by stars.  Points are slightly offset
  along the $x$-axis in both cases in order to make it easier to
  distinguish between the different calibrations.
  \label{fig:te}}
\end{figure}

Examples of the differences in the effective area for the different
calibrations are shown in Figure~\ref{fig:ea} for both ACIS-I
(CL0016+1609; top) and ACIS-S (MS1054.5-0321; bottom) observations of
galaxy clusters.  Shown are the effective areas (left) and fractional
residuals from the 4.2 calibration (right) for the 3.1 (red), 4.1
(blue), and 4.2 (black) calibration versions.  The vertical dotted
lines denote the 0.7-7.0 keV energy range used in this analysis.
Residuals are defined as $\Delta A/A^{4.2}=(A^{x}-A^{4.2})/A^{4.2}$,
where $x$ refers to 3.1 or 4.1.  The same definition of residual
applies to the other cluster properties with $A$ replaced by the
parameter of interest.  The resulting temperatures follow the same
trends as the normalization of the effective areas ($T_e \propto
A(E)$).

The results from the Markov chain analysis are summarized in
Table~\ref{tab:xspec} for all three calibration versions.  The basic
trend is as expected from the calibration notes and the effective area
curves, $T_e^{3.1} > T_e^{4.2} > T_e^{4.1}$.  A2163 appears in the
table twice.  Following the methodology outlined above results in what
seems like unrealistically high temperatures, $16 - 20$ keV between
the calibrations.  In addition, the combination of its formally small
uncertainties and it being very far from the mean inferred Hubble
parameter (due to its high $T_e$) causes undue influence on the
resultant Hubble parameter (see Section~\ref{sec:H0}).  This cluster
is among a handful of clusters that are known to have \ion{H}{1}
column densities that are significantly different from the
\citet{dickey1990} values \citep[e.g.,][]{govoni2004}.  When adopting
the updated column density of $18.7 \times 10^{20}$ cm$^{-2}$
\citep{govoni2004} instead of the \citet{dickey1990} value of $12.1
\times 10^{20}$ cm$^{-2}$ the temperatures and other parameters change
significantly because the column density strongly affects the lower
energy part of the spectrum.  The updated temperatures are in the
range $10 - 14$ keV, placing A2163 more in line with previous
results and closer to the mean Hubble constant from the other
clusters, lessening the impact from this single cluster.  As an aside,
also note that a newer \ion{H}{1} column density study yields an even lower
value of $10.9 \times 10^{20}$ cm$^{-2}$ \citep{kalberla2005}.
Choosing an updated \ion{H}{1} column density destroys our uniform, systematic
study by treating one cluster special.  It is tempting to throw out
this cluster to keep the uniformity.  We therefore present results for
both the \citet{dickey1990} and \citet{govoni2004} $N_{\mathrm{H}}$ values and
also explore the effects of removing A2163 from the sample.  Initially
A2163 has a large impact on the final $H_0$ value.  However, once the
new value of $N_{\mathrm{H}}$ is adopted, A2163 basically has no effect on the
results.  Table~\ref{tab:xspec} shows the results for both column
density values.  The figures show results for only the adopted updated
$N_{\mathrm{H}}$ value and a star is used to denote A2163 to distinguish it from
the other clusters.

R02 adopted $N_{\mathrm{H}}$ values from spectral fits where available (five
clusters), used the value from the Bell Labs \ion{H}{1} survey
\citep{stark1992} as adopted in a detailed analysis for that one
cluster \citep{donahue1999}, and adopted the DL values
\citep{dickey1990} for the remaining clusters.  In particular, the
adopted value for A2163 is $N_{\mathrm{H}}=16.5 \times 10^{20}$ cm$^{-2}$
\citep{elbaz1995}.  A quick spectral fit to the \chandra\ data
including $N_{\mathrm{H}}$ as a free parameter suggests $N_{\mathrm{H}} \sim 16 \times
10^{20}$ cm$^{-2}$.  Further investigation is beyond the scope of this
work.  We are concerned with the overall differences between
calibrations more so than the values of the individual derived
quantities for a particular cluster.

Figure~\ref{fig:te} shows the comparison of the temperatures from the
version 3.1 (red) and 4.1 (blue) calibration versus that of the 4.2
calibration (left) along with the fractional residuals compared to the
4.2 calibration results (right).  Also plotted are the \asca\ results
(green) for the 17 overlapping clusters used for SZE/X-ray distances
using the same SZE data as B06 but using \rosat\ and \asca\ data
(R02).\nocite{reese2002} The dotted line shows the one-to-one
correspondence.  There is a clear division between the 3.1 and 4.1
results on either side of the equality line, with 3.1 falling above
the relation and 4.1 falling below, clearly seen in the residuals.
A2163 results are shown with stars.  

\begin{figure}[!tbh]
\centerline{
  \includegraphics[width=85mm]{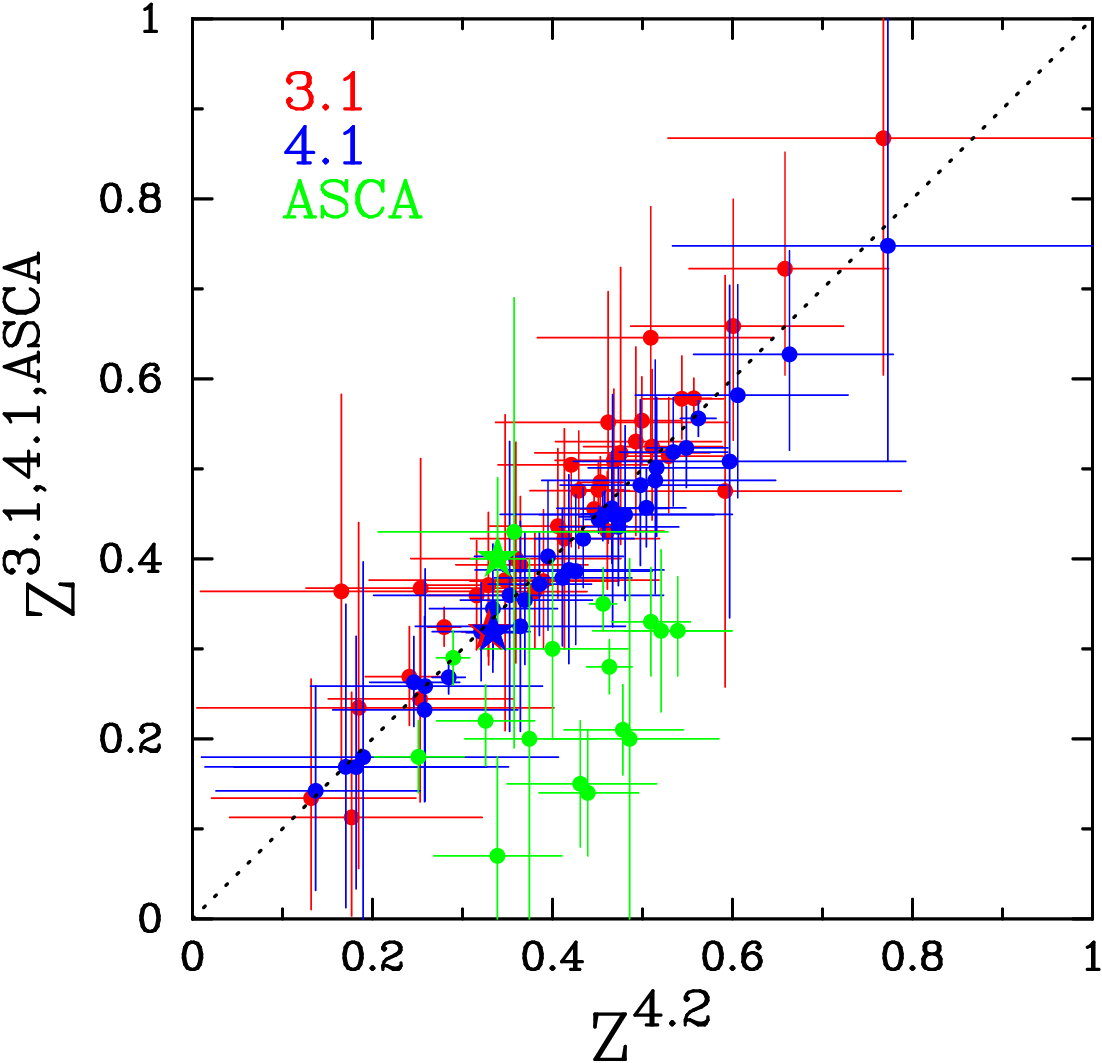}
  \includegraphics[width=85mm]{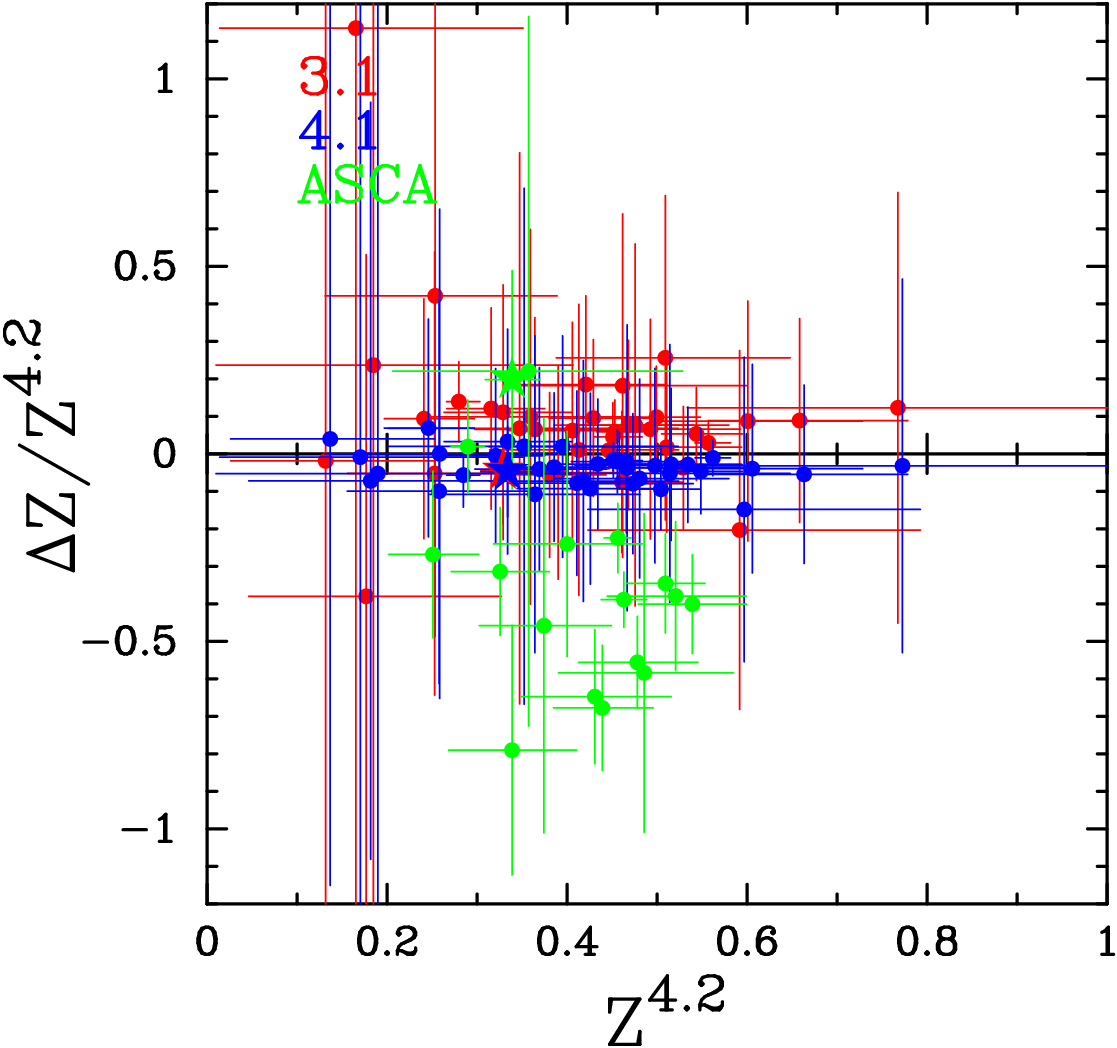}
}
\caption{Comparison of different abundance measurements (left) and the
  corresponding fractional residuals from the 4.2 calibration results
  (right).  Abundance determinations using the 3.1 calibration (red)
  and 4.1 (blue) calibration are compared against metallicities using
  the newest 4.2 version.  Also plotted are the \asca\ (green)
  abundances adopted in R02
  for comparison of the 17 clusters that overlap with this work. The
  dotted black line shows the equality relation.  Error bars show 68\%
  confidence statistical uncertainties and the A2163 results are
  denoted by stars.  Points are slightly offset along the $x$-axis in
  both cases in order to make it easier to distinguish between the
  different calibrations.
  \label{fig:abund}}
\end{figure}

The mean ratios between temperatures are
\begin{eqnarray}
  \left < \frac{T_e^{3.1}}{T_e^{4.2}} \right > = 1.06 \pm 0.05, \ \
  \left < \frac{T_e^{4.1}}{T_e^{4.2}} \right > = 0.93 \pm 0.03, \ 
\label{eq:te_ratio}
\end{eqnarray}
where the above uncertainty is simply the rms in the ratio.  The
changes between the two older calibrations and the newest are roughly
the same order ($\sim 6$\%) but in different directions.  Therefore,
the average change in temperatures between the 3.1 and 4.1
calibrations is of order $\sim 12$\%, which would produce a $\sim
24$\% difference in the inferred Hubble parameter.

\begin{figure}[!tbh]
\centerline{
  \includegraphics[width=85mm]{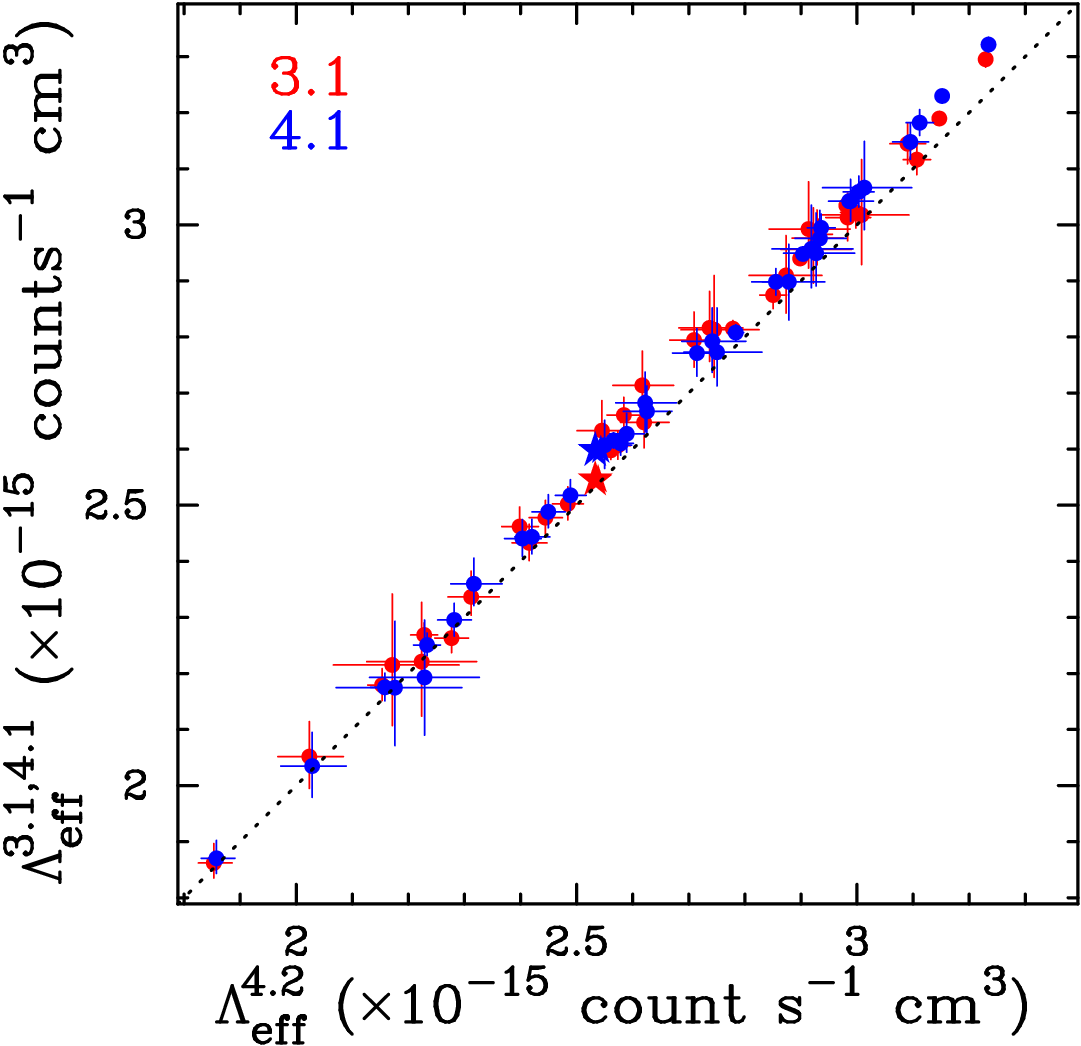}
  \includegraphics[width=85mm]{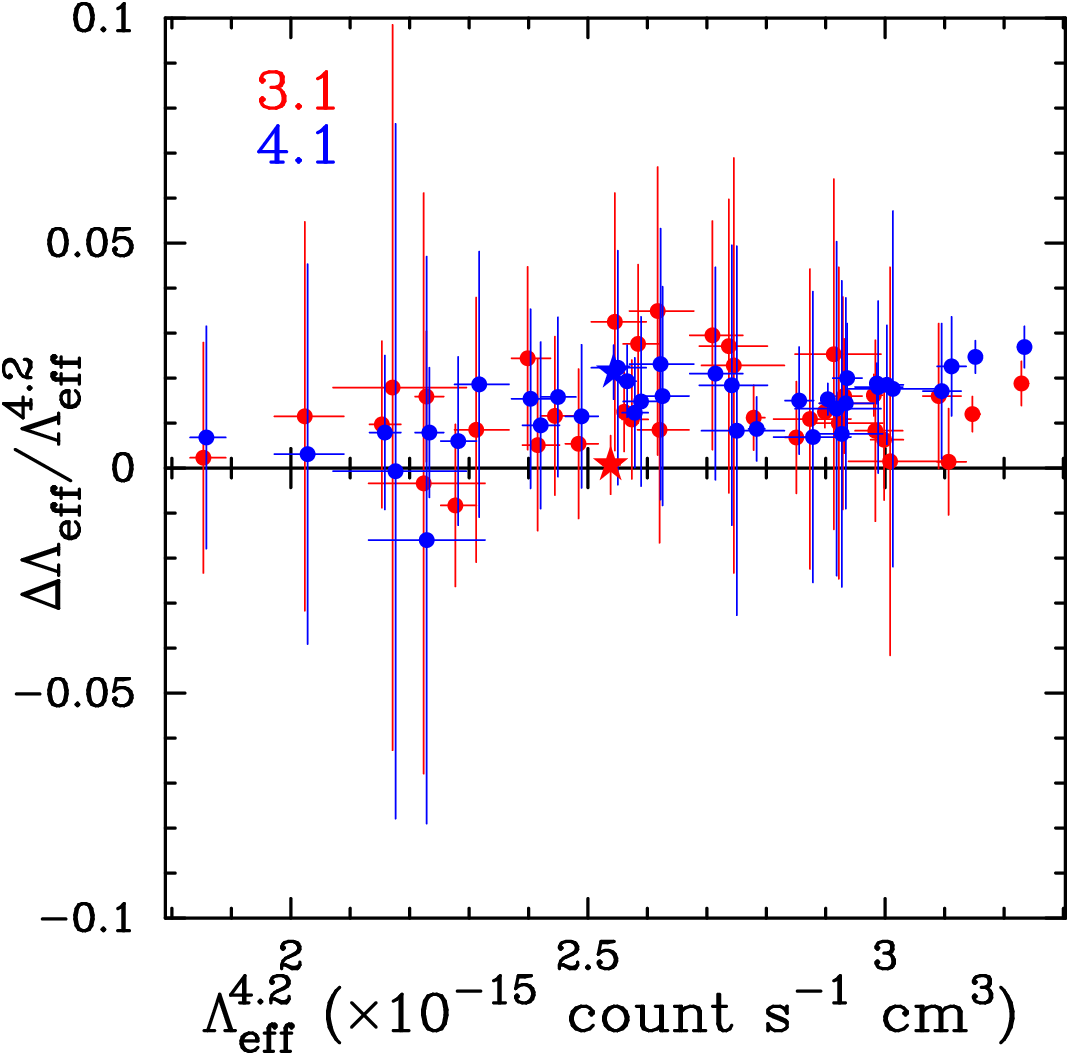}
}
\caption{Comparison of different effective X-ray cooling functions
  (left) and the corresponding fractional residuals (right).
  Effective cooling function determinations using calibration versions
  3.1 (red) and 4.1 (blue) are compared against emissivities using the
  new 4.2 calibration results.  The dotted black line shows the
  equality relation.  Error bars show 68\% confidence statistical
  uncertainties and the A2163 results are denoted by stars.  Points
  are slightly offset along the $x$-axis in both cases in order to make
  it easier to distinguish between the different calibrations.
  \label{fig:emiss}}
\end{figure}

\begin{deluxetable*}{lcccccc}
\tablewidth{0pt}
\tablecolumns{6}
\tablecaption{Compilation of Mean Ratios: Updated A2163 $N_{\mathrm{H}}$ \label{tab:ratios}}
\tablehead{
\colhead{Parameter} & \colhead{3.1/4.2} & \colhead{4.1/4.2} &
\colhead{3.1/B06} & \colhead{4.1/B06} & \colhead{4.2/B06} & \colhead{\asca/4.2}
}
\startdata
$T_e$          & $1.06 \pm 0.05$ & $0.93 \pm 0.03$ & $1.05 \pm 0.11$ 
               & $0.92 \pm 0.10$ & $0.99 \pm 0.11$
               & $0.98 \pm 0.12$\\
$Z$            & $1.08 \pm 0.21$ & $0.96 \pm 0.04$ & $1.16 \pm 0.43$ 
               & $1.03 \pm 0.31$ & $1.08 \pm 0.34$
               & $0.66 \pm 0.28$\\
$\Lambda_{\rm eff}$ & $1.01 \pm 0.01$ & $1.01 \pm 0.01$ & $1.03 \pm 0.07$ 
               & $1.03 \pm 0.07$ & $1.02 \pm 0.07$
               & \nodata \\
$f_{(\nu,T_e)}$ & $0.998 \pm 0.002$ & $1.002\pm 0.001$ & $0.999 \pm 0.003$ 
               & $1.003 \pm 0.004$ & $1.001 \pm 0.003$
               & \nodata \\
$A(1\mbox{kev})$\tablenotemark{a} & $1.01 \pm 0.02$ & $0.95 \pm 0.01$ & $0.96 \pm 0.10 $
               & $0.91 \pm 0.10$ & $0.95 \pm 0.10$ 
               & \nodata \\
$d_A$\tablenotemark{b} & $0.93 \pm 0.08$ & $1.13 \pm 0.06$ & $1.06 \pm 0.24$ 
               & $1.29 \pm 0.29$ & $1.15 \pm 0.25$
               & $1.07 \pm 0.37$\\
\enddata
\tablenotetext{a}{B06 are the effective areas from the 3.1 calibration
  using only those data sets that appear in \citealt{bonamente2006}.}
\tablenotetext{b}{B06 are the published distances from \citealt{bonamente2006}.}
\end{deluxetable*}

The mean ratios comparing against the B06 \nocite{bonamente2006} results are
\begin{eqnarray}
  \left < \frac{T_e^{3.1}}{T_e^{B06}} \right > = 1.05 \pm 0.11, \ \
  \left < \frac{T_e^{4.1}}{T_e^{B06}} \right > = 0.92 \pm 0.10, \ \
  \left < \frac{T_e^{4.2}}{T_e^{B06}} \right > = 0.99 \pm 0.11.
\label{eq:te_ratio_bon}
\end{eqnarray}
Therefore it is likely that there is a small ($\sim 5$\%) overall
systematic between the spectral analysis of B06 \nocite{bonamente2006}
and this work.  Given the various decisions on background regions,
spectral extraction regions, and so forth, this is reasonable
agreement.  This analysis removes those uncertainties between analyses
by performing the same systematic analysis on the same data for three
different \chandra\ calibration versions.  We focus on the differences
between the calibrations rather than the values themselves.  The mean
ratios of the temperatures as well as other parameters considered here
are summarized in Table~\ref{tab:ratios}.

The abundances from the spectral fits (left) and the fractional
residuals with the 4.2 results (right) are plotted in
Figure~\ref{fig:abund}.  Again calibration versions 3.1 (red) and 4.1
(blue) are plotted against the 4.2 calibration and the dotted line
shows equality.  Also shown are the \asca\ abundances adopted in R02
\nocite{reese2002}.  There is no clear trend in the \chandra\ results
as is the case for the temperatures.  There is a mild offset between
the \chandra\ and \asca\ metallicities.  The \asca\ results are
compiled from the literature and the data are not uniformly analyzed,
the abundances used for the analyses often differing.  The abundances
do change the effective cooling functions, $\Lambda_{\rm eff}$.
However, the effects of metallicity on $\Lambda_{\rm eff}$ are small,
typically on the order of $\sim 1$\% and $\lesssim 5$\% even for
changing the abundance by factors of 2 or 3.

Derived effective cooling functions from the spectral analysis (left)
and corresponding fractional residuals from the 4.2 calibration
(right) are illustrated in Figure~\ref{fig:emiss}.  Results for
calibration versions 3.1 (red) and 4.1 (blue) are shown against that of
version 4.2.  The equality relation is also shown.  There is a clear
trend for $\Lambda_{\rm eff}$ to be greater in 3.1 and 4.1 compared to
the 4.2 results but as shown by the residuals, it is a small effect,
$<4$\% for 3.1 and $<2$\% for the 4.1 results.

The abundance of metals does not directly enter the distance
calculation.  However, it indirectly enters through the X-ray cooling
function, $\Lambda_{\rm eff}$.  The X-ray temperature also indirectly
enters the calculation through relativistic corrections to the
frequency dependence of the SZE, $f_{(\nu, T_e)}$, where $\nu$ is the
frequency of the observations (see Section~\ref{sec:H0}).  Mean ratios
and rms's for the temperature, abundance, X-ray cooling function, SZE
frequency function including relativistic corrections
\citep{itoh1998}, the effective area at 1~keV, and angular diameter
distance (see Section~\ref{sec:H0}) are summarized in
Table~\ref{tab:ratios}.

\section{Implications to the estimate of $H_0$}  
\label{sec:H0}

Direct angular diameter distances, $d_A$, from a combined SZE and
X-ray analysis are straight forward to compute, in theory, \citep[see,
  for example][]{birkinshaw1991, reese2002, bonamente2006},
particularly for the simple isothermal $\beta$-model.  The actual
observations and parameter extraction from the data are not as
straight forward.  We concern ourselves here with only the components
of the distance calculation that involve X-ray spectral properties or
are derived from them.  SZE/X-ray derived distances have the following
dependence on X-ray spectral properties
\begin{equation}
d_A \propto \frac{\Lambda_{\rm eff}}{T_e^2 f^2_{(\nu, T_e)} S_X}
\propto \frac{\Lambda_{\rm eff} A(E_{\rm fid})}{T_e^2 f^2_{(\nu, T_e)}},
\label{eq:da_xspec_depend}
\end{equation}
where $T_e$ is the electron temperature, $\Lambda_{\rm eff}$ is the
effective X-ray cooling function (see Section~\ref{subsec:det_to_cgs}), and
$f_{(\nu, T_e)}$ is the spectral dependence of the SZE at frequency,
$\nu$, including relativistic corrections, which depend on $T_e$
\citep[e.g.,][]{itoh1998, challinor1998}, $S_X$ is the X-ray surface
brightness that depends on the effective area that changes between the
calibrations, $A(E)$ is the effective area of the observatory, and we
have used the fact that $S_X \propto {\cal N}_X/A(E_{\rm fid})$, where
${\cal N}_X$ is the number of observed counts and is constant.  Since
the exposure maps are all computed at 1 keV, $S_X$ simply changes by
the ratio of effective areas at $E_{\rm fid}= 1$~keV.  It is not
immediately obvious how the spectral results will affect the final
Hubble constant due to the complexity of the dependencies on spectral
parameters.  However, it is clear that a $\sim 10$\% change in $T_e$
will have an appreciable effect on the distances and therefore on the
inferred Hubble constant.

Cluster spatial properties from $\beta$-model fits (B06)
\nocite{bonamente2006} are adopted for the distance calculation.
X-ray spectral properties from this analysis for the three
calibrations are combined with the adopted cluster spatial properties
in order to compute distances to each galaxy cluster.
We estimate the uncertainty on $d_A$ by backing out the X-ray spectral
variable uncertainty from the published uncertainty and including the
new \chandra\ calibration spectral result uncertainties assuming
everything adds in quadrature.  Namely we compute the uncertainty in
$d_A$ by
\begin{equation}
\left ( \frac{\delta d_A^{x}}{d_A^{x}} \right ) ^2 =
\left ( \frac{\delta d_A^{B06}}{d_A^{B06}} \right ) ^2 
- 2 \left ( \frac{\delta T_e^{B06}}{T_e^{B06}} \right ) ^2
- \left ( \frac{\delta \Lambda_{\rm eff}^{B06}}{\Lambda_{\rm eff}^{B06}} \right ) ^2
+ 2 \left ( \frac{\delta T_e^{x}}{T_e^{x}} \right ) ^2
+ \left ( \frac{\delta \Lambda_{\rm eff}^{x}}{\Lambda_{\rm eff}^{x}} \right ) ^2,
\label{eq:da_uncertainty}
\end{equation}
where $x$ refers to one of the three calibrations of this work and
$B06$ refers to the published values (B06). \nocite{bonamente2006}  The
average of the positive and negative uncertainties is used for the
uncertainty.  This method preserves, as best we can, the correlations
among the parameters.  In particular, $r_c$ and $\beta$ from the
$\beta$-model are strongly correlated \citep[for an example in this
  context see][]{reese2000}.  However, the spectral results are
independent of the $\beta$-model.  Therefore we can preserve this
correlation through uncertainty propagation with this method.  The
temperature uncertainty comes with a factor of 2 because the angular
diameter distance is inversely proportional to temperature squared,
$d_A \propto T_e^{-2}$.  We include the additional sources of
statistical uncertainty, $\sim 19$\%, from Table~3 of
B06 \nocite{bonamente2006} by adding in quadrature to the uncertainties
computed from Equation~(\ref{eq:da_uncertainty}).  These total
statistical uncertainties are then used for the $H_0$ calculation and are
summarized in Table~\ref{tab:xspec}, which shows the derived angular
diameter distances with 68\% statistical uncertainties.

The Hubble constant is computed by performing a $\chi^2$ fit to the
cluster distances using the theoretical angular diameter distance
relation for a flat, $\Lambda$-dominated universe with $\Omega_\Lambda
= 0.73$ and $\Omega_m = 0.27$ consistent with the {\it WMAP} results
\citep{komatsu2010, komatsu2009}.  The resulting angular diameter
distances for the full sample and updated $N_{\mathrm{H}}$ for A2163
yield
\begin{eqnarray}
H_0^{3.1} &=& 70.0 \pm 3.7 \mbox{ km  s$^{-1}$ Mpc$^{-1}$}, \\
H_0^{4.1} &=& 55.4 \pm 2.9 \mbox{ km  s$^{-1}$ Mpc$^{-1}$}, \\
H_0^{4.2} &=& 63.7 \pm 3.3 \mbox{ km  s$^{-1}$ Mpc$^{-1}$}, 
\label{eq:h0_3calibs}
\end{eqnarray}
where the uncertainties are statistical only at 68\% confidence with
$\chi^2 = 40.6$, $34.4$, $38.8$ for 37 degrees of freedom for the 3.1,
4.1, and 4.2 calibrations respectively.  Solely from the changing
calibration, taking the 4.2 calibration as the baseline,
$H_0^{3.1}/H_0^{4.2} = 1.10$ and $H_0^{4.1}/H_0^{4.2}=0.87$, showing
changes of roughly 10\% and 13\% in the determination of the Hubble
constant, respectively.  There is a $\sim 23$\% change in $H_0$
between the 3.1 and 4.1 calibrations, consistent with the expectation
from the mean of the ratio of temperatures (see
Section~\ref{sub:spec_results} and Table~\ref{tab:ratios}).

\begin{figure}[!tbh]
  \centerline{
    \includegraphics[width=85mm]{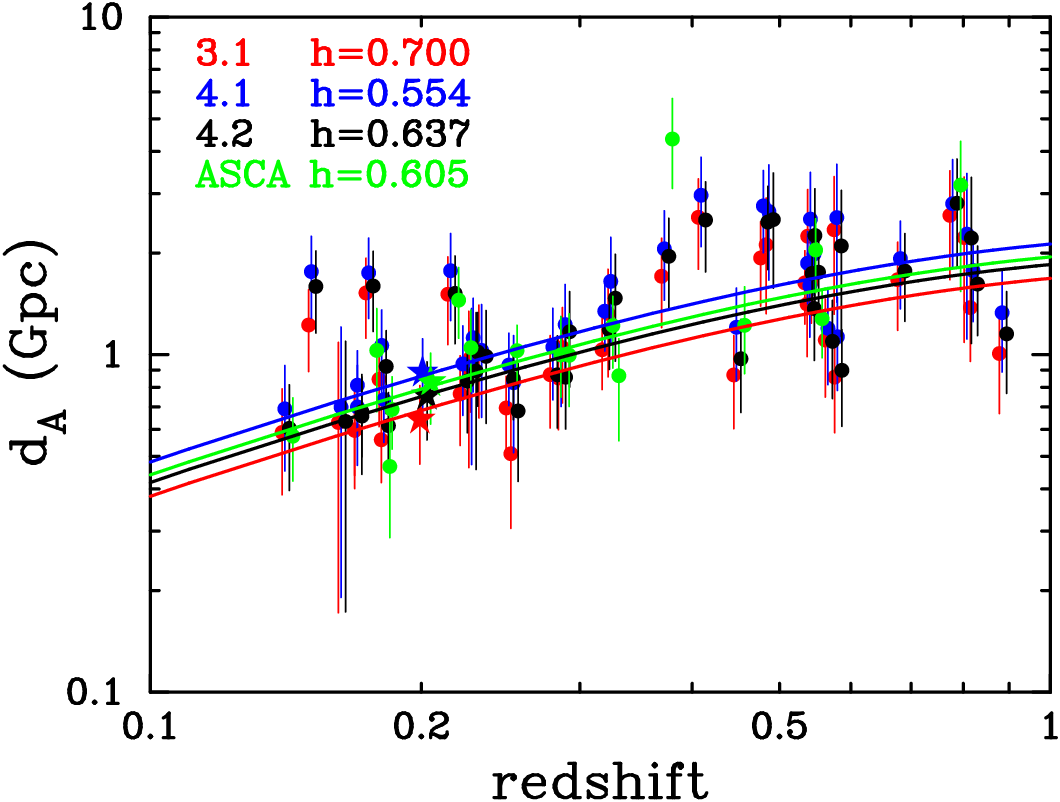}
    \includegraphics[width=85mm]{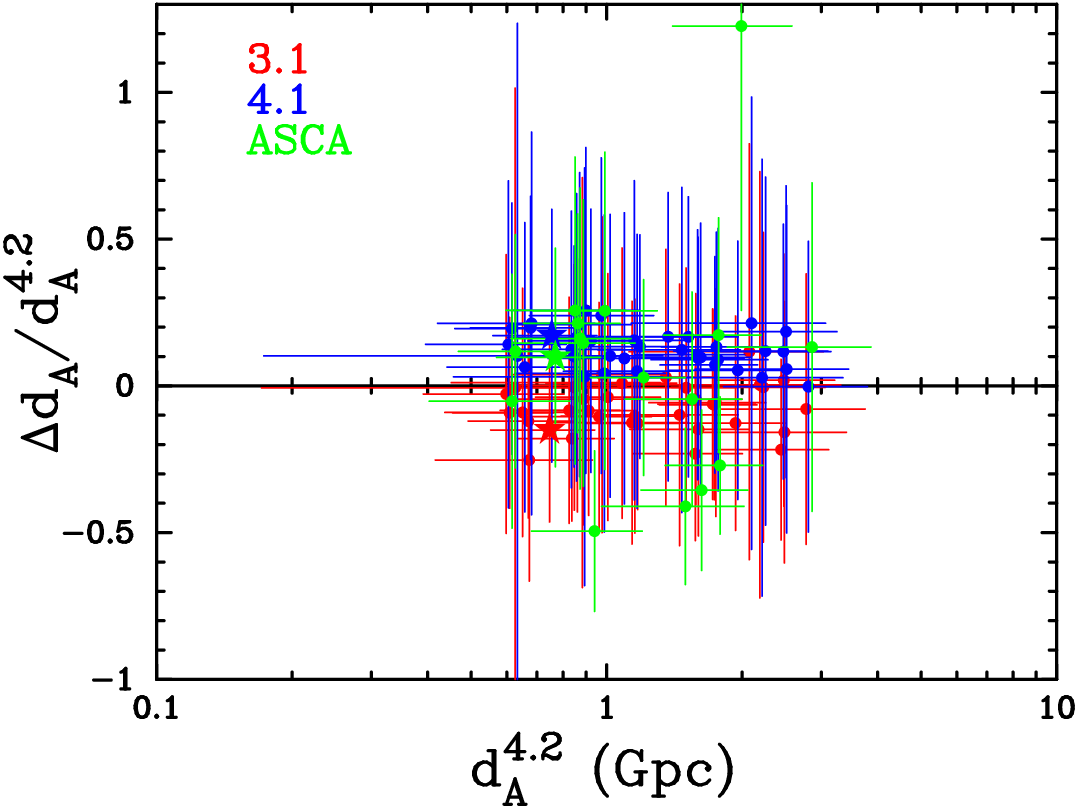}
  }
  \centerline{
    \includegraphics[width=85mm]{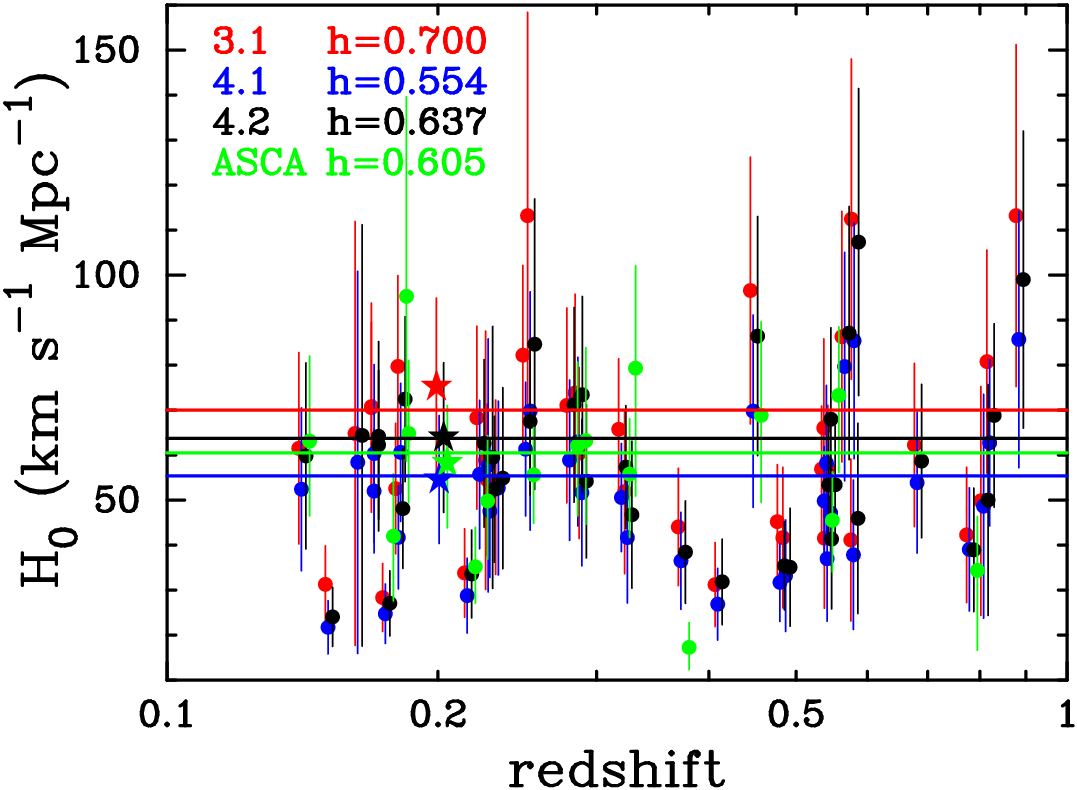}
    \includegraphics[width=85mm]{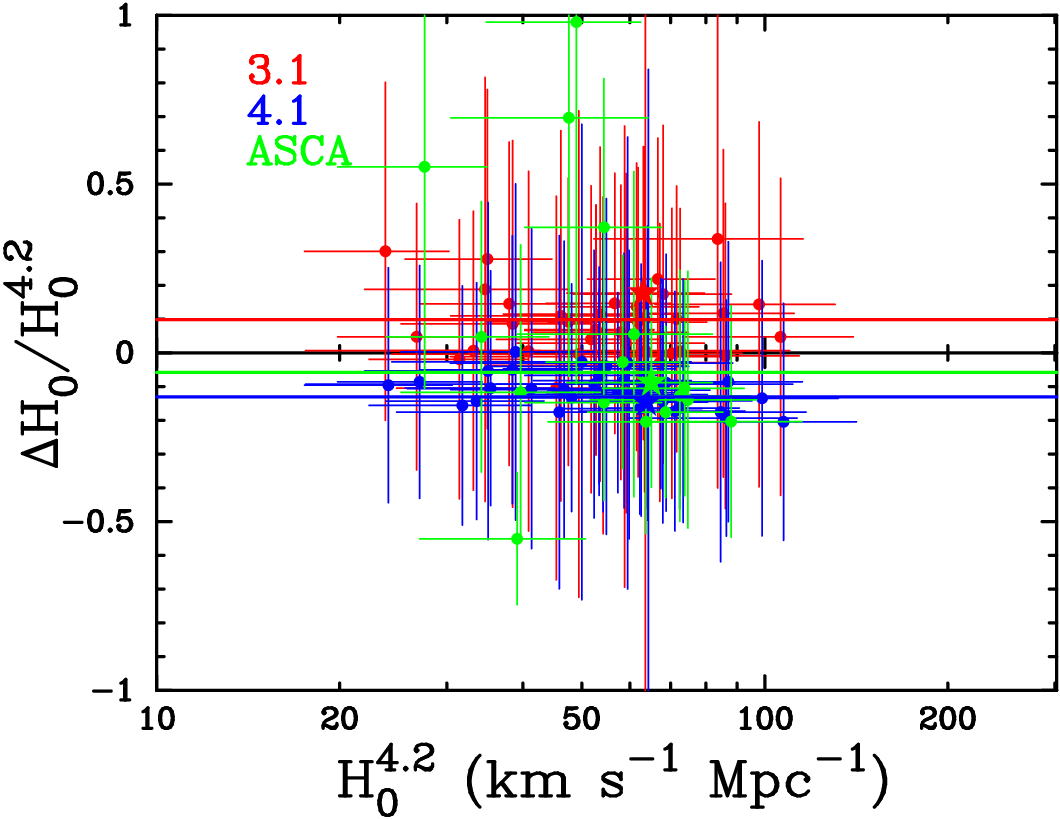}
  }
  \caption{Angular diameter distances (top) and corresponding Hubble
    parameters (bottom) determined using the results from the 3.1 (red),
    4.1 (blue), and 4.2 (black) calibrations.  In both cases, the right
    panels show the fractional residuals from the 4.2 calibration
    results. Results from R02 
    are also plotted for comparison and denoted as \asca\ (green).
    The best0fit theoretical angular diameter distance relations are
    also shown for each calibration (top, left) with corresponding
    Hubble parameters in the legend.  The best-fit Hubble parameters
    are shown in the $H_0$ figure (bottom, left) and corresponding
    residuals (bottom, right).  Error bars show 68\% confidence statistical
    uncertainties and the A2163 results are denoted by stars.  Points
    are offset slightly along the $x$-axis in order to make it easier to
    distinguish between the different calibrations. \label{fig:da_z}}
\end{figure}

In Figure~\ref{fig:da_z}, we plot the angular diameter distances (top)
and corresponding Hubble constants (bottom) determined using the
results from the 3.1 (red), 4.1 (blue), and 4.2 (black) calibrations
for each cluster.  In both cases, the right panels show the fractional
residuals from the 4.2 calibration results. Results from R02
\nocite{reese2002} are also plotted for comparison and denoted as
\asca\ (green).  The best-fit theoretical angular diameter distance
relations (top) and corresponding Hubble constants (bottom) are also
shown.  Angular diameter distances are also summarized in
Table~\ref{tab:xspec} for each of the calibrations.  Uncertainties are
68\% confidence and include both the uncertainty from the fit and the
additional sources of statistical uncertainty from Table~3 of B06
\nocite{bonamente2006}.  In general, the distances follow the trends
expected from the temperatures, namely, $d_A^{3.1} < d_A^{4.2} <
d_A^{4.1}$, indicating that the temperature changes dominate the
changes in distances.

Hubble parameters are computed from the determined angular diameter
distances using $\chi^2$ for the original $N_{\mathrm{H}}=12.1 \times 10^{20}$
cm$^{-2}$ for A2163, the updated A2163 $N_{\mathrm{H}}$, and excluding A2163 from
the fit for each of the three calibrations.  The results are compiled
in Table~\ref{tab:h0_results}, where the uncertainties are 68\%
confidence statistical uncertainties and the number in parentheses is
the $\chi^2$ at the best-fit value.  Hubble parameters are computed
from $\chi^2$ fits for the 17 clusters that overlap in the R02
\nocite{reese2002} sample, becoming 16 clusters when A2163 is
excluded.  A simple average is also computed for each case by first
converting the $d_A$ for each cluster into a Hubble parameter and then
averaging.  Finally, a $\chi^2$ fit is done for the published values
of B06 and R02 both including and excluding A2163.

Table~\ref{tab:h0_results} summarizes the full set of Hubble parameter
estimates.  The beginning of each section marks the $H_0$ estimation
method ($\chi^2$ or average) followed by the number of clusters used
in each calculation.  The last two sections are the results of
performing $\chi^2$ fits both including and not including A2163 using
the published distances from B06 and R02.  In the case of R02, the
first entry includes the entire sample of 18 clusters although only 17
of which overlap with the B06 sample.  A fit of just those 17 appears
in the second column and then the results without A2163 in the third
column.

\begin{deluxetable*}{lccc}
\tablewidth{0pt}
\tablecolumns{4}
\tablecaption{Compilation of $H_0$ Results \label{tab:h0_results}}
\tablehead{
\colhead{$H_0$} & \colhead{DL A2163 $N_{\mathrm{H}}$} & \colhead{Updated A2163 $N_{\mathrm{H}}$} & \colhead{No A2163} 
}
\startdata
$\chi^2$ full sample & 38 & 38 & 37 \\
$H_0^{3.1}$   & $82.8 \pm 4.6 (75.9)$ & $70.0 \pm 3.7 (40.6)$ & $69.7 \pm 3.7 (40.5)$ \\
$H_0^{4.1}$   & $58.4 \pm 3.1 (42.4)$ & $55.4 \pm 2.9 (34.3)$ & $55.5 \pm 2.9 (34.3)$ \\
$H_0^{4.2}$   & $68.8 \pm 3.7 (52.2)$ & $63.7 \pm 3.3 (38.8)$ & $63.7 \pm 3.4 (38.8)$ \\
\hline
$\chi^2$ R02 overlap & 17 & 17 & 16\\
$H_0^{3.1}$   & $90.1 \pm 7.0 (48.5)$ & $66.9 \pm 4.7 (15.8)$ & $66.1 \pm 4.8 (15.5)$ \\
$H_0^{4.1}$   & $58.2 \pm 4.1 (21.2)$ & $52.3 \pm 3.6 (11.8)$ & $52.2 \pm 3.8 (11.8)$ \\
$H_0^{4.2}$   & $70.1 \pm 5.1 (28.7)$ & $60.5 \pm 4.3 (14.4)$ & $60.2 \pm 4.4 (14.3)$ \\
\hline
Avg full sample      & 38 & 38 & 37\\
$H_0^{3.1}$    & $66.1 \pm 30.3$ & $62.9 \pm 21.7$ & $62.6 \pm 21.9$ \\
$H_0^{4.1}$    & $52.5 \pm 17.2$ & $51.3 \pm 15.4$ & $51.2 \pm 15.6$ \\
$H_0^{4.2}$    & $59.7 \pm 22.1$ & $58.0 \pm 18.9$ & $57.8 \pm 19.1$ \\
\hline
Avg R02 overlap     & 17 & 17 & 16\\
$H_0^{3.1}$    & $68.3 \pm 36.1$ & $61.2 \pm 17.6$ & $60.4 \pm 17.7$ \\
$H_0^{4.1}$    & $51.8 \pm 16.9$ & $49.2 \pm 12.0$ & $48.9 \pm 12.3$ \\
$H_0^{4.2}$    & $60.0 \pm 23.3$ & $56.1 \pm 15.5$ & $55.6 \pm 15.9$ \\
\hline
\hline
$\chi^2$ B06 refit & 38 && 37\\
$H_0^{B06}$       & $76.2 \pm 4.1 (55.9)$ & \nodata & $73.5 \pm 4.1 (51.7)$\\
\hline
$\chi^2$ R02 refit & 18 & 17 & 16\\
$H_0^{R02}$  & $60.8 \pm 4.0 (16.5)$ & $60.5 \pm 4.1 (16.4)$ & $60.7 \pm 4.3 (16.4)$ \\
\enddata
\tablecomments{DL is the \citet{dickey1990} \ion{H}{1} survey.  The first row
  of each section describes the type of $H_0$ calculation ($\chi^2$ or
  average) followed by the number of clusters used in the fit for each
  of the three cases considered.  The numbers in parentheses are the
  $\chi^2$ value at the best fit.  The last two sections are fits to
  the published B06 and R02 distances.  For R02, the first column is
  the full sample of R02, the second column include the 17 overlapping
  clusters with this work, and the third column then also excludes
  A2163 from the $\chi^2$ fit.}
\end{deluxetable*}

Using the means of the ratios of the X-ray spectral parameters with
respect to the calibration version 4.2 results
(Table~\ref{tab:ratios}) and scaling $H_0^{4.2}$ implies
$H_0^{3.1}=69.2$ km s$^{-1}$ Mpc$^{-1}$ and $H_0^{3.1}=56.5$ km
s$^{-1}$ Mpc$^{-1}$, very close to the Hubble parameters from the
$\chi^2$ analysis.  This is also true when dropping A2163 from the
analysis but not true when using the \citet{dickey1990} $N_{\mathrm{H}}$ value
for A2163.  In that case, using the ratios to scale the 4.2 results
predicts $H_0^{3.1}=75.5$ km s$^{-1}$ Mpc$^{-1}$ and $H_0^{3.1}=62.3$
km s$^{-1}$ Mpc$^{-1}$ compared to the $\chi^2$ results 82.8 and 58.4
km s$^{-1}$ Mpc$^{-1}$, respectively.

\section{Discussion}
\label{sec:discussion}

We perform a uniform, systematic spectral analysis of 38 galaxy
clusters using three different \chandra\ calibrations and find
significant differences in the inferred spectral properties of galaxy
clusters between the calibrations.  Using the newest calibration, 4.2,
as the baseline to which to compare, the temperatures change $\sim
6$\% on average in the 3.1 and 4.1 calibrations, but in opposite
directions.  In particular, the temperature changes between the
extreme cases (3.1 and 4.1) show a $\sim 13$\% difference in $T_e$, on
average.  These results are consistent with an analysis of 10 galaxy
clusters that found the 4.1 calibration results yield $\sim 10$\%
lower temperatures than the 3.1 calibration \citep{gaetz2009}.  The
differences in spectral properties of galaxy clusters between the
calibrations may have profound implications on inferred cosmological
parameters from galaxy cluster studies.

Using the simple isothermal $\beta$-model as a vehicle we explore the
ramifications of the effects of the \chandra\ calibration on a
particular cosmological application, distances to galaxy clusters from
a combined analysis of SZE and X-ray data.  Because $H_0 \propto
T_e^2$, changes in cluster temperatures have a potentially large
impact on the inferred Hubble parameter.  The 3.1, 4.1, and 4.2
calibrations imply Hubble constants of 70.0, 55.4, and 63.7 km
s$^{-1}$ Mpc$^{-1}$ using the updated column density for A2163
results.  The results remain essentially unchanged when removing A2163
from the sample to preserve the uniformity of the analysis to the last
detail.  Simply from the effects of the \chandra\ calibration, the 3.1
results yield a 10\% higher $H_0$ than the most recent 4.2 calibration
and the 4.1 results yield a 13\% lower $H_0$ than the 4.2 calibration
results.  This is in rough agreement with what the mean ratios of
temperatures would predict, 13\% and 14\% effects for the 3.1 and 4.1
calibrations, respectively.  This strongly suggests that, although
there are a number of quantities that change with the spectral
results, the change in $T_e$ is, by far, the most important when
considering SZE/X-ray derived distance based estimates of the Hubble
parameter.  In the most extreme case, comparing the 3.1 to the 4.1
results, there is a $\sim 24$\% change in the Hubble parameter due to
the change in \chandra\ calibration.

Although the isothermal assumption is over-simplistic, it is
sufficient to study the effects of the new calibration on SZE/X-ray
derived distances.  B06 \nocite{bonamente2006} showed that the Hubble
parameter estimates from SZE/X-ray distances do not depend as strongly
on the model as one might naively believe.  In particular, the
isothermal $\beta$-model yields results consistent with the more
sophisticated hydrostatic equilibrium model and an isothermal
$\beta$-model with the central regions of the X-ray data removed that
they considered.  This simple isothermal $\beta$-model also
facilitates isolating the effects that the different spectral
parameters have on the distances and therefore on the Hubble constant.

SZE/X-ray distances tend to favor a Hubble constant of order 60 km
s$^{-1}$ Mpc$^{-1}$ \citep[e.g.,][]{carlstrom2002, reese2002} with a
few exceptions \citep[e.g.,][]{bonamente2006, mason2001}.  This is in
contrast to other probes that favor $H_0 \sim 70$ km s$^{-1}$
Mpc$^{-1}$, such as the {\it HST} $H_0$ key project
\citep{freedman2001}, recent supernova results
\citep[e.g.,][]{hicken2009}, and recent {\it WMAP} results
\citep{komatsu2009, komatsu2010}.  Potential systematics in SZE/X-ray
derived estimates of $H_0$ are still formidable \citep[see for
  example, ][]{birkinshaw1991, hughes1998, reese2002, bonamente2006},
making the results consistent within the uncertainties.  It is still
curious that SZE/X-ray distances tend to favor a lower Hubble
constant.

One possible explanation of this bias in $H_0$ from SZE/X-ray
determined distances to galaxy clusters is the combined effects from
the degree of inhomogeneity and the multi-temperature structure of the
ICM \citep{kawahara2008}.  The bias vanishes in the
limit of an isothermal and homogeneous ICM, an idealized and
unrealistic limit.  The presence of inhomogeneity in the cluster gas
is suggested by a detailed study of the nearby galaxy cluster A3667
\citep{kawahara2008b}, which finds 30\%$-$40\% rms density fluctuations
in that cluster.  In addition, 20\% rms temperature fluctuations have
been seen in Hydra A \citep{simionescu2009}.  The theoretical
underpinning of the inhomogeneity model was boosted because the nearby
galaxy cluster A3667 exhibits the expected lognormal signature
\citep{kawahara2008b}.

This model implies that the bias in the Hubble constant can be
decomposed into three factors
\begin{equation}
\label{eq:f_h}
f_H \equiv \frac{H_{\rm 0,est}}{H_{\rm 0,true}} = \chi_{\sigma} \chi_{T}(T_{\rm
  ew}) \frac{\chi_{T}(T_{\rm spec})}{\chi_{T}(T_{\rm ew})}.
\end{equation}
where $\chi_{\sigma}$, $\chi_{T}(T_{\rm ew})$, and $\chi_{T}(T_{\rm
  spec})/\chi_{T}(T_{\rm ew})$ represent the systematic errors due to
the presence of ICM gas inhomogeneities, non-isothermality, and the
difference between the spectroscopic ($T_{\rm spec}$), and
emission-weighted ($T_{\rm ew}$) temperatures.  Numerical values for
each of the above bias factors, of course, depend crucially on the
degree of inhomogeneities and the temperature structure of the ICM.
However, simulated clusters suggest that these biases are roughly ($10
- 30$)\% overestimate, ($0 - 20$)\% underestimate, and ($10 - 20$)\%
underestimate, respectively, resulting in an overall ($10 - 20$)\%
underestimate in the Hubble parameter.  For analytic expressions and
further details, see \citet{kawahara2008}.

We note that the above result is consistent with the many studies of
the statistical and systematic uncertainties in Hubble constant
determinations from SZE/X-ray distances \citep[e.g.,][]{inagaki1995,
  kobayashi1996, yoshikawa1998, hughes1998, sulkanen1999, wang2006}
that find only small biases if any at all.  All of these studies use
an emission-weighted temperature for the simulated clusters.  This
corresponds to neglecting the third factor in Equation~(\ref{eq:f_h}),
and thus leading to no substantial bias because the first and second
factors coincidentally compensate each other.  \citet{mazzotta2004}
are the first to point out clearly that the spectroscopic temperature,
$\Tspec$, is systematically lower than the emission-weighted
temperature, $\Tew$ \citep[see also][]{mathiesen2001, rasia2005}, so
that the third factor in Equation~(\ref{eq:f_h}) is essential
\citep{kawahara2007, kawahara2008}.

This bias can theoretically be accounted for with more realistic
models of the ICM.  For example, once the variance of the ICM
inhomogeneity is known, a fit to even a simplistic model for the
temperature profile largely avoids the bias in $H_0$ from SZE/X-ray
distances \citep{kawahara2008}.  Progress on this front will require a
multi-wavelength approach combining recent SZE experiments, deep X-ray
observations, and weak lensing measurements.


The large potential systematics in SZE/X-ray derived distances and
inferred Hubble constant means that the various results are consistent
with each other and other probes of $H_0$ \citep[for detail of the
  systematics, see, e.g.,][]{reese2002, bonamente2006}.  Here, we took a
detailed look at one of those systematics, the effect of the
\chandra\ calibration on cluster temperatures.  There have been very
little work on including calibration uncertainties in X-ray analysis.
However, a Monte Carlo approach to incorporate calibration uncertainty
for parameter estimation from \chandra\ ACIS-S observations has been
developed \citep{drake2006}.  Better methods of incorporating
uncertainties, especially systematics such as instrumental
calibration, will need to be developed.

There are alternative methods to X-ray spectroscopy for determining
galaxy cluster temperatures.  Theoretical studies of non-parametric
deprojection methods of SZE and X-ray imaging data on both idealized
and simulated clusters suggest that cluster temperature profiles may
be accurately reconstructed without X-ray spectroscopy
\citep{yoshikawa1999, puchwein2006, ameglio2007}.  Both parametric and
non-parametric methods applied to actual SZE and X-ray data also show
broad agreement with temperature profiles derived from X-ray
spectroscopy \citep{kitayama2004, nord2009, mroczkowski2009}.  Cluster
temperatures may also be inferred from SZE data only if one assumes the
value for the gas mass fraction of the cluster \citep{laroque2006,
  joy2001}.  Because these methods use only X-ray imaging data, they
do not require the longer exposure times necessary for spectroscopic
$T_e$ measurements.  In addition, these alternative temperature
measurements may, in theory, alleviate some of the dependence of
derived cosmological parameters on the \chandra\ calibration.
However, most methods currently have large uncertainties on the
cluster temperature and are complicated by the fact that they still
depend on the \chandra\ (or other X-ray observatory) calibration
through the effective area for both exposure maps and the cooling
function calculation.

The determination of galaxy cluster temperatures is particularly
important because it is widely used to infer the gravitational mass of
clusters.  Potential systematics on cluster temperatures will have a
strong impact on cosmological parameters when using clusters as probes
of cosmology.  In particular, temperature systematics will strongly
affect the normalization of the matter power spectrum, $\sigma_8$,
through the cluster temperature$-$halo mass relation in conjunction with
cluster abundances \citep[e.g.,][]{rasia2005, shimizu2006}.

Analysis of observations of galaxy clusters does have the potential to
provide insight into cosmology and has been successful in the past,
favoring low $\Omega_M$ long before hints of a cosmological constant
appeared \citep[e.g.,][]{white1993}.  Surveys of galaxy clusters have
the tantalizing appeal that they probe the growth of structure, one of
the few probes to do so, and have the potential to constrain tightly
the equation of state of the dark energy \citep[e.g.,][]{bartlett1994,
  holder2000, haiman2001, majumdar2004}.  However, in order to realize
that potential, both galaxy clusters themselves and the details of the
instruments must be understood precisely which will be feasible with
combined efforts from ongoing observations and planned missions and
observatories covering a wide range of wavelengths.

\acknowledgements

We are grateful to Akio Hoshino and Yoshitaka Ishisaki for useful
discussions about the \chandra\ calibration that helped to motivate this
work.  We thank John Carlstrom, Massimiliano Bonamente, and Marshall
Joy for a careful reading and comments on a draft of this manuscript.
We are indebted to the \chandra\ help desk and calibration teams for
answering many questions about the details of the
\chandra\ calibration. This research has made extensive use of data
obtained from the \chandra\ Data Archive and software provided by the
\chandra\ X-ray Center (CXC).  E.D.R gratefully acknowledges support
from a JSPS Postdoctoral Fellowship for Foreign Researches (P07030).
Y.S.  acknowledges the support from the Global Scholars Program of
Princeton University, and thanks people in Peyton Hall, Princeton
University, for their warm hospitality and discussions.  This work is
supported by Grant-in-Aid for Scientific research of Japanese Ministry
of Education, Culture, Sports, Science and Technology (Nos.  18740112,
18072002, 20$\cdot$08324, 20$\cdot$10466, 20340041, 22$\cdot$5467, and
21740139), and by JSPS (Japan Society for Promotion of Science)
Core-to-Core Program ``International Research Network for Dark
Energy''.


{
}


\clearpage
\clearpage
\LongTables
\begin{deluxetable*}{lccccrcr}
\tablewidth{0pt}
\tablecolumns{8}
\tabletypesize{\tiny}
\tablecaption{\chandra\ Data \label{tab:data}}
\tablehead{
	\colhead{} &
	\colhead{R.A.} &
	\colhead{Decl.} &
	\colhead{} &
	\colhead{$N_{\mathrm{H}}$\tablenotemark{a}} &
	\colhead{} &
	\colhead{} &
	\colhead{Livetime}
\\
	\colhead{Cluster} &
	\colhead{(h m s)} &
	\colhead{(d m s)} &
	\colhead{$z$} &
	\colhead{($\times 10^{20}$ cm$^{-2}$)} &
	\colhead{ObsID} &
	\colhead{Array} &
	\colhead{(ks)}
}
\startdata
CL0016$+$1609	   &$00\phn 18\phn 33.5$&$+16\phn 26\phn 12.5$
		   &0.541& 4.07 
		   &520	  &I&   66\\
A0068		   &$00\phn 37\phn 06.2$&$+09\phn 09\phn 33.2$
                   &0.255& 4.94 
		   &3250  &I&  10\\
A0267		   &$01\phn 52\phn 42.1$&$+01\phn 00\phn 35.7$
		   &0.230& 2.80 
	           &1448  &I&  8\\
               &&&&&3580  &I&  20\\
A0370		   &$02\phn 39\phn 53.2$&$-01\phn 34\phn 35.0$
                   &0.375& 3.06
                   &515	  &S&  66\\
               &&&&&7715  &I&   7\\
MS0451.6$-$0305	   &$04\phn 54\phn 11.4$&$-03\phn 00\phn 52.7$
                   &0.550& 5.03
                   &529	  &I&  14\\
               &&&&&902	  &S&  43\\
MACSJ0647.7$+$7015   &$06\phn 47\phn 50.2$&$+70\phn 14\phn 54.6$
                   &0.584& 5.63
                   &3196  &I&  19\\
               &&&&&3584  &I&  20\\
A0586		   &$07\phn 32\phn 20.2$&$+31\phn 37\phn 55.6$
                   &0.171& 5.15
                   &530	  &I&  10\\
MACSJ0744.8$+$3927   &$07\phn 44\phn 52.8$&$+39\phn 27\phn 26.7$
                   &0.686& 5.68
                   &3197  &I&  20\\
               &&&&&3585  &I&  19\\
               &&&&&6111  &I&  49\\
A0611  	           &$08\phn 00\phn 56.6$&$+36\phn 03\phn 24.1$
                   &0.288& 4.99
                   &3194  &S&  36\\
A0665		   &$08\phn 30\phn 58.1$&$+65\phn 50\phn 51.6$	  
                   &0.182& 4.24 
                   &531	  &I&   9\\
               &&&&&3586  &I&  30\\
               &&&&&7700  &I&   5\\
A0697		   &$08\phn 42\phn 57.5$&$+36\phn 21\phn 56.2$	  
                   &0.282& 3.41 
                   &532	  &I&   8\\
               &&&&&4217  &I&  19\\
A0773		   &$09\phn 17\phn 52.8$&$+51\phn 43\phn 38.9$	  
                   &0.217& 1.44 
                   &533	  &I&  11\\
               &&&&&3588  &I&   9\\
               &&&&&5006  &I&  20\\
Zwicky 3146	   &$10\phn 23\phn 39.7$&$+04\phn 11\phn 09.5$	  
                   &0.291& 3.01 
                   &909	  &I&  46\\
               &&&&&9371  &I&  40\\
MS1054.5$-$0321	   &$10\phn 56\phn 59.4$&$-03\phn 37\phn 34.2$	  
                   &0.826& 3.58 
                   &512	  &S&  84\\
MS1137.5$+$6625	   &$11\phn 40\phn 22.3$&$+66\phn 08\phn 16.0$	  
                   &0.784& 1.21 
                   &536	  &I& 117\\
MACSJ1149.5$+$2223 &$11\phn 49\phn 35.5$&$+22\phn 24\phn 02.3$	 
                   &0.544& 2.28 
                   &1656  &I&  18\\
               &&&&&3589  &I&  20\\
A1413		   &$11\phn 55\phn 18.0$&$+23\phn 24\phn 17.0$	  
                   &0.142& 2.19 
                   &537	  &I&  10\\
               &&&&&1661  &I&  10\\
               &&&&&5002  &I&  36\\
               &&&&&5003  &I&  75\\
               &&&&&7696  &I&   5\\
CLJ1226.9$+$3332     &$12\phn 26\phn 57.9$&$+33\phn 32\phn 47.4$	  
                   &0.890& 1.38 
                   &932	  &S&  10\\
               &&&&&3180  &I&  32\\
               &&&&&5014  &I&  32\\
MACSJ1311.0$-$0310 &$13\phn 11\phn 01.7$&$-03\phn 10\phn 38.5$	 
                   &0.490& 1.88 
                   &3258  &I&  15\\
               &&&&&6110  &I&  63\\
               &&&&&7721  &I&   7\\
               &&&&&9381  &I&  30\\
A1689		   &$13\phn 11\phn 29.5$&$-01\phn 20\phn 28.2$	  
                   &0.183& 1.82 
                   &540	  &I&  10\\
               &&&&&1663  &I&  11\\
               &&&&&5004  &I&  20\\
               &&&&&6930  &I&  76\\
               &&&&&7289  &I&  75\\
               &&&&&7701  &I&   5\\
RXJ1347.5$-$1145   &$13\phn 47\phn 30.6$&$-11\phn 45\phn 08.6$	  
                   &0.451& 4.85 
                   &506	  &S&   9\\
               &&&&&507	  &S&  10\\
               &&&&&3592  &I&  57\\
MS1358.4$+$6245	   &$13\phn 59\phn 50.6$&$+62\phn 31\phn 04.1$	  
                   &0.327& 1.93 
                   &516	  &S&  51\\
               &&&&&7714  &I&   7\\
A1835		   &$14\phn 01\phn 02.0$&$+02\phn 52\phn 41.7$	  
                   &0.252& 2.32 
                   &495	  &S&  19\\
               &&&&&496	  &S&  11\\
               &&&&&6880  &I& 117\\
               &&&&&6881  &I&  36\\
               &&&&&7370  &I&  39\\
MACSJ1423.8$+$2404   &$14\phn 23\phn 47.9$&$+24\phn 04\phn 42.6$	 
                   &0.545& 2.83 
                   &1657  &I&  18\\
               &&&&&4195  &S& 115\\
A1914		   &$14\phn 26\phn 00.8$&$+37\phn 49\phn 35.7$	  
                   &0.171& 0.95 
                   &542	  &I&   8\\
               &&&&&3593  &I&  19\\
A1995		   &$14\phn 52\phn 57.9$&$+58\phn 02\phn 55.8$	  
                   &0.322& 1.42 
                   &906	  &S&  57\\
               &&&&&7021  &I&  48\\
               &&&&&7713  &I&   7\\
A2111		   &$15\phn 39\phn 41.0$&$+34\phn 25\phn 08.8$	  
                   &0.229& 1.93 
                   &544	  &I&  10\\
A2163		   &$16\phn 15\phn 46.2$&$-06\phn 08\phn 51.3$	  
                   &0.202& 12.1/18.7\tablenotemark{b}
                   &545	  &I&   9\\
               &&&&&1653  &I&  71\\

A2204		   &$16\phn 32\phn 46.9$&$+05\phn 34\phn 31.9$	  
                   &0.152& 5.67 
                   &499	  &S&  10\\
               &&&&&6104  &I&  10\\
               &&&&&7940  &I&  77\\
A2218		   &$16\phn 35\phn 51.9$&$+66\phn 12\phn 34.5$	  
                   &0.176& 3.24 
                   &553	  &I&   6\\
               &&&&&1454  &I&  11\\
               &&&&&1666  &I&  44\\
               &&&&&7698  &I&   5\\
RXJ1716.4$+$6708   &$17\phn 16\phn 48.8$&$+67\phn 12\phn 34.5$	  
                   &0.813& 3.70 
                   &548	  &I&  52\\
A2259		   &$17\phn 20\phn 08.5$&$+27\phn 40\phn 11.0$	 
                   &0.164& 3.70 
                   &3245  &I&  10\\
A2261		   &$17\phn 22\phn 27.1$&$+32\phn 07\phn 57.8$	  
                   &0.224& 3.28 
                   &550	  &I&   9\\
               &&&&&5007  &I&  24\\
MS2053.7$-$0449	   &$20\phn 56\phn 21.2$&$-04\phn 37\phn 47.8$	  
                   &0.583& 4.96 
                   &551	  &I&  44\\
               &&&&&1667  &I&  44\\
MACSJ2129.4$-$0741 &$21\phn 29\phn 26.0$&$-07\phn 41\phn 28.7$	 
                   &0.570& 4.84 
                   &3199  &I&  18\\
               &&&&&3595  &I&  19\\
RXJ2129.7$+$0005   &$21\phn 29\phn 39.9$&$+00\phn 05\phn 19.8$	  
                   &0.235& 4.28 
                   &552	  &I&  10\\
MACSJ2214.9$-$1359 &$22\phn 14\phn 57.3$&$-14\phn 00\phn 12.3$	 
                   &0.483& 3.28 
                   &3259  &I&  19\\
               &&&&&5011  &I&  18\\
MACSJ2228.5$+$2036 &$22\phn 28\phn 33.0$&$+20\phn 37\phn 14.4$	 
                   &0.412& 4.58 
                   &3285  &I&  20\\
\enddata
\tablenotetext{a}{\citet{dickey1990} values.}
\tablenotetext{b}{Both values are considered because A2163 is known to
have an $N_{\mathrm{H}}$ significantly different from \citet{dickey1990}.}
\end{deluxetable*}

\clearpage
\tabletypesize{\tiny}
\LongTables
\begin{landscape}
  \begin{deluxetable*}{lrccccrccccrccc}
\tablewidth{0pt}
\tablecolumns{15}
\tabletypesize{\tiny}
\tablecaption{Spectroscopic Results \label{tab:xspec}}
\tablehead{
  \colhead{} & \multicolumn{4}{c}{3.1} && \multicolumn{4}{c}{4.1} && \multicolumn{4}{c}{4.2}\\
  \cline{2-5}\cline{7-10}\cline{12-15}
  \colhead{} & \colhead{$T_e$} & \colhead{$Z$} &  \colhead{$\Lambda_{\rm eff}$\tablenotemark{a}} & \colhead{$d_A$} & \colhead{} & \colhead{$T_e$} & \colhead{$Z$} & \colhead{$\Lambda_{\rm eff}$\tablenotemark{a}} & \colhead{$d_A$} & \colhead{} &\colhead{$T_e$} & \colhead{$Z$} & \colhead{$\Lambda_{\rm eff}$\tablenotemark{a}} & \colhead{$d_A$}\\
  \colhead{} & \colhead{(keV)} & \colhead{($Z_\sun$)} & \colhead{} & \colhead{(Gpc)} & \colhead{} & \colhead{(keV)} & \colhead{($Z_\sun$)} & \colhead{} & \colhead{(Gpc)} & \colhead{} & \colhead{(keV)} & \colhead{($Z_\sun$)} & \colhead{} & \colhead{(Gpc)}
}
\startdata
CL0016$+$1609        & $ 9.93^{+0.51}_{-0.47} $ & $ 0.371^{+0.081}_{-0.079} $ & $ 2.502^{+0.030}_{-0.028} $ & $ 1.632^{+0.400}_{-0.400} $ && $ 8.99^{+0.39}_{-0.36} $ & $ 0.345^{+0.067}_{-0.065} $ & $ 2.518^{+0.028}_{-0.026} $ & $ 1.864^{+0.451}_{-0.451} $ && $ 9.39^{+0.49}_{-0.41} $ & $ 0.334^{+0.072}_{-0.070} $ & $ 2.489^{+0.028}_{-0.027} $ & $ 1.740^{+0.425}_{-0.425} $  \\
A0068              & $11.28^{+1.64}_{-1.19} $ & $ 0.475^{+0.239}_{-0.218} $ & $ 3.018^{+0.099}_{-0.088} $ & $ 0.509^{+0.203}_{-0.203} $ && $ 8.57^{+0.82}_{-0.74} $ & $ 0.508^{+0.176}_{-0.160} $ & $ 3.066^{+0.082}_{-0.075} $ & $ 0.825^{+0.313}_{-0.313} $ && $ 9.66^{+0.99}_{-0.85} $ & $ 0.597^{+0.196}_{-0.174} $ & $ 3.013^{+0.084}_{-0.074} $ & $ 0.681^{+0.259}_{-0.259} $  \\
A0267              & $ 6.91^{+0.32}_{-0.30} $ & $ 0.530^{+0.105}_{-0.104} $ & $ 2.976^{+0.050}_{-0.048} $ & $ 0.978^{+0.303}_{-0.303} $ && $ 6.24^{+0.29}_{-0.27} $ & $ 0.482^{+0.090}_{-0.086} $ & $ 2.976^{+0.050}_{-0.037} $ & $ 1.122^{+0.347}_{-0.347} $ && $ 6.67^{+0.30}_{-0.30} $ & $ 0.498^{+0.095}_{-0.089} $ & $ 2.934^{+0.047}_{-0.044} $ & $ 1.018^{+0.315}_{-0.315} $  \\
A0370              & $ 8.78^{+0.42}_{-0.40} $ & $ 0.375^{+0.079}_{-0.075} $ & $ 2.263^{+0.028}_{-0.026} $ & $ 1.707^{+0.504}_{-0.504} $ && $ 7.90^{+0.37}_{-0.34} $ & $ 0.403^{+0.080}_{-0.079} $ & $ 2.295^{+0.030}_{-0.029} $ & $ 2.058^{+0.608}_{-0.608} $ && $ 8.14^{+0.39}_{-0.38} $ & $ 0.395^{+0.084}_{-0.082} $ & $ 2.282^{+0.030}_{-0.029} $ & $ 1.955^{+0.578}_{-0.578} $  \\
MS0451.6$-$0305      & $10.78^{+0.55}_{-0.53} $ & $ 0.504^{+0.094}_{-0.091} $ & $ 2.179^{+0.029}_{-0.028} $ & $ 1.641^{+0.401}_{-0.401} $ && $ 9.59^{+0.50}_{-0.43} $ & $ 0.386^{+0.076}_{-0.074} $ & $ 2.175^{+0.025}_{-0.024} $ & $ 1.986^{+0.484}_{-0.484} $ && $10.37^{+0.54}_{-0.52} $ & $ 0.426^{+0.085}_{-0.081} $ & $ 2.158^{+0.027}_{-0.026} $ & $ 1.753^{+0.430}_{-0.430} $  \\
MACSJ0647.7$+$7015   & $12.07^{+1.22}_{-1.00} $ & $ 0.367^{+0.144}_{-0.138} $ & $ 2.633^{+0.053}_{-0.049} $ & $ 0.858^{+0.271}_{-0.271} $ && $10.06^{+0.80}_{-0.69} $ & $ 0.259^{+0.107}_{-0.105} $ & $ 2.607^{+0.045}_{-0.042} $ & $ 1.130^{+0.346}_{-0.346} $ && $11.53^{+1.24}_{-0.94} $ & $ 0.259^{+0.130}_{-0.127} $ & $ 2.550^{+0.048}_{-0.044} $ & $ 0.899^{+0.285}_{-0.285} $  \\
A0586              & $ 7.16^{+0.43}_{-0.39} $ & $ 0.659^{+0.141}_{-0.127} $ & $ 2.956^{+0.074}_{-0.060} $ & $ 0.598^{+0.197}_{-0.197} $ && $ 6.39^{+0.34}_{-0.32} $ & $ 0.582^{+0.120}_{-0.110} $ & $ 2.949^{+0.072}_{-0.058} $ & $ 0.700^{+0.229}_{-0.229} $ && $ 6.67^{+0.36}_{-0.35} $ & $ 0.606^{+0.122}_{-0.114} $ & $ 2.927^{+0.068}_{-0.058} $ & $ 0.658^{+0.216}_{-0.216} $  \\
MACSJ0744.8$+$3927   & $ 8.99^{+0.44}_{-0.39} $ & $ 0.436^{+0.086}_{-0.080} $ & $ 2.462^{+0.035}_{-0.033} $ & $ 1.667^{+0.485}_{-0.485} $ && $ 8.03^{+0.36}_{-0.34} $ & $ 0.379^{+0.072}_{-0.068} $ & $ 2.441^{+0.034}_{-0.032} $ & $ 1.926^{+0.559}_{-0.559} $ && $ 8.57^{+0.37}_{-0.37} $ & $ 0.411^{+0.077}_{-0.075} $ & $ 2.403^{+0.033}_{-0.032} $ & $ 1.768^{+0.513}_{-0.513} $  \\
A0611              & $ 6.89^{+0.27}_{-0.27} $ & $ 0.393^{+0.076}_{-0.073} $ & $ 2.478^{+0.031}_{-0.029} $ & $ 0.852^{+0.254}_{-0.254} $ && $ 6.28^{+0.25}_{-0.24} $ & $ 0.354^{+0.069}_{-0.063} $ & $ 2.488^{+0.031}_{-0.028} $ & $ 0.999^{+0.297}_{-0.297} $ && $ 6.93^{+0.27}_{-0.27} $ & $ 0.369^{+0.075}_{-0.072} $ & $ 2.449^{+0.030}_{-0.029} $ & $ 0.857^{+0.255}_{-0.255} $  \\
A0665              & $ 8.47^{+0.27}_{-0.28} $ & $ 0.359^{+0.061}_{-0.061} $ & $ 2.983^{+0.027}_{-0.026} $ & $ 0.845^{+0.233}_{-0.233} $ && $ 7.30^{+0.19}_{-0.18} $ & $ 0.319^{+0.051}_{-0.050} $ & $ 2.995^{+0.025}_{-0.024} $ & $ 1.064^{+0.292}_{-0.292} $ && $ 7.96^{+0.26}_{-0.25} $ & $ 0.321^{+0.054}_{-0.054} $ & $ 2.936^{+0.025}_{-0.024} $ & $ 0.923^{+0.254}_{-0.254} $  \\
A0697              & $10.50^{+0.59}_{-0.57} $ & $ 0.518^{+0.206}_{-0.101} $ & $ 3.013^{+0.044}_{-0.042} $ & $ 0.872^{+0.266}_{-0.266} $ && $ 9.26^{+0.50}_{-0.41} $ & $ 0.449^{+0.088}_{-0.083} $ & $ 3.042^{+0.039}_{-0.036} $ & $ 1.052^{+0.318}_{-0.318} $ && $10.32^{+0.56}_{-0.55} $ & $ 0.481^{+0.099}_{-0.095} $ & $ 2.988^{+0.041}_{-0.039} $ & $ 0.871^{+0.265}_{-0.265} $  \\
A0773              & $ 8.13^{+0.32}_{-0.32} $ & $ 0.509^{+0.079}_{-0.076} $ & $ 3.145^{+0.037}_{-0.036} $ & $ 1.509^{+0.438}_{-0.438} $ && $ 7.25^{+0.24}_{-0.21} $ & $ 0.435^{+0.063}_{-0.060} $ & $ 3.148^{+0.033}_{-0.031} $ & $ 1.772^{+0.511}_{-0.511} $ && $ 7.99^{+0.32}_{-0.31} $ & $ 0.473^{+0.067}_{-0.065} $ & $ 3.095^{+0.032}_{-0.031} $ & $ 1.519^{+0.441}_{-0.441} $  \\
ZW3146             & $ 6.74^{+0.10}_{-0.10} $ & $ 0.477^{+0.030}_{-0.029} $ & $ 2.815^{+0.014}_{-0.014} $ & $ 1.048^{+0.329}_{-0.329} $ && $ 6.00^{+0.10}_{-0.09} $ & $ 0.447^{+0.027}_{-0.026} $ & $ 2.808^{+0.014}_{-0.014} $ & $ 1.227^{+0.385}_{-0.385} $ && $ 6.21^{+0.09}_{-0.09} $ & $ 0.456^{+0.028}_{-0.026} $ & $ 2.784^{+0.014}_{-0.014} $ & $ 1.169^{+0.367}_{-0.367} $  \\
MS1054.5$-$0321      & $11.30^{+1.30}_{-0.99} $ & $ 0.134^{+0.132}_{-0.124} $ & $ 1.862^{+0.034}_{-0.027} $ & $ 1.378^{+0.423}_{-0.423} $ && $ 9.79^{+0.84}_{-0.75} $ & $ 0.142^{+0.109}_{-0.105} $ & $ 1.871^{+0.032}_{-0.027} $ & $ 1.774^{+0.522}_{-0.522} $ && $10.39^{+0.92}_{-0.83} $ & $ 0.137^{+0.116}_{-0.111} $ & $ 1.858^{+0.033}_{-0.027} $ & $ 1.618^{+0.479}_{-0.479} $  \\
MS1137.5$+$6625      & $ 6.54^{+0.64}_{-0.55} $ & $ 0.376^{+0.184}_{-0.167} $ & $ 2.052^{+0.062}_{-0.057} $ & $ 2.586^{+0.919}_{-0.919} $ && $ 6.05^{+0.53}_{-0.45} $ & $ 0.359^{+0.168}_{-0.153} $ & $ 2.035^{+0.060}_{-0.056} $ & $ 2.800^{+0.981}_{-0.981} $ && $ 6.04^{+0.56}_{-0.46} $ & $ 0.352^{+0.171}_{-0.151} $ & $ 2.028^{+0.061}_{-0.056} $ & $ 2.808^{+0.989}_{-0.989} $  \\
MACSJ1149.5$+$2223   & $ 9.80^{+0.77}_{-0.68} $ & $ 0.244^{+0.117}_{-0.115} $ & $ 2.648^{+0.048}_{-0.045} $ & $ 1.410^{+0.423}_{-0.423} $ && $ 8.92^{+0.63}_{-0.55} $ & $ 0.232^{+0.094}_{-0.094} $ & $ 2.667^{+0.045}_{-0.043} $ & $ 1.600^{+0.474}_{-0.474} $ && $ 9.85^{+0.74}_{-0.67} $ & $ 0.258^{+0.103}_{-0.102} $ & $ 2.625^{+0.044}_{-0.042} $ & $ 1.370^{+0.409}_{-0.409} $  \\
A1413              & $ 7.65^{+0.12}_{-0.11} $ & $ 0.485^{+0.028}_{-0.028} $ & $ 3.295^{+0.014}_{-0.014} $ & $ 0.588^{+0.203}_{-0.203} $ && $ 6.86^{+0.08}_{-0.08} $ & $ 0.450^{+0.024}_{-0.023} $ & $ 3.321^{+0.013}_{-0.013} $ & $ 0.690^{+0.238}_{-0.238} $ && $ 7.46^{+0.09}_{-0.08} $ & $ 0.458^{+0.025}_{-0.025} $ & $ 3.234^{+0.007}_{-0.007} $ & $ 0.605^{+0.209}_{-0.209} $  \\
CLJ1226.9$+$3332     & $14.18^{+1.33}_{-1.26} $ & $ 0.113^{+0.139}_{-0.110} $ & $ 2.337^{+0.046}_{-0.033} $ & $ 1.007^{+0.338}_{-0.338} $ && $11.95^{+1.14}_{-0.91} $ & $ 0.169^{+0.125}_{-0.123} $ & $ 2.360^{+0.046}_{-0.039} $ & $ 1.331^{+0.442}_{-0.442} $ && $13.10^{+1.23}_{-1.06} $ & $ 0.182^{+0.145}_{-0.135} $ & $ 2.317^{+0.050}_{-0.041} $ & $ 1.152^{+0.384}_{-0.384} $  \\
MACSJ1311.0$-$0310   & $ 6.50^{+0.26}_{-0.24} $ & $ 0.431^{+0.066}_{-0.065} $ & $ 2.661^{+0.032}_{-0.032} $ & $ 2.115^{+0.792}_{-0.792} $ && $ 5.55^{+0.17}_{-0.16} $ & $ 0.456^{+0.064}_{-0.059} $ & $ 2.627^{+0.036}_{-0.033} $ & $ 2.655^{+0.991}_{-0.991} $ && $ 5.82^{+0.19}_{-0.16} $ & $ 0.466^{+0.063}_{-0.061} $ & $ 2.589^{+0.032}_{-0.030} $ & $ 2.512^{+0.938}_{-0.938} $  \\
A1689              & $11.55^{+0.12}_{-0.12} $ & $ 0.324^{+0.022}_{-0.021} $ & $ 3.189^{+0.009}_{-0.009} $ & $ 0.559^{+0.141}_{-0.141} $ && $ 9.76^{+0.12}_{-0.11} $ & $ 0.268^{+0.017}_{-0.016} $ & $ 3.230^{+0.008}_{-0.008} $ & $ 0.736^{+0.186}_{-0.186} $ && $10.81^{+0.12}_{-0.12} $ & $ 0.285^{+0.018}_{-0.018} $ & $ 3.152^{+0.008}_{-0.008} $ & $ 0.616^{+0.156}_{-0.156} $  \\
RXJ1347.5$-$1145     & $14.35^{+0.39}_{-0.39} $ & $ 0.553^{+0.049}_{-0.048} $ & $ 2.598^{+0.017}_{-0.016} $ & $ 0.870^{+0.267}_{-0.267} $ && $11.81^{+0.29}_{-0.23} $ & $ 0.457^{+0.038}_{-0.037} $ & $ 2.615^{+0.014}_{-0.013} $ & $ 1.205^{+0.369}_{-0.369} $ && $13.36^{+0.39}_{-0.38} $ & $ 0.504^{+0.044}_{-0.043} $ & $ 2.566^{+0.015}_{-0.015} $ & $ 0.972^{+0.299}_{-0.299} $  \\
MS1358.4$+$6245      & $ 7.45^{+0.36}_{-0.29} $ & $ 0.525^{+0.085}_{-0.082} $ & $ 2.433^{+0.033}_{-0.032} $ & $ 1.324^{+0.464}_{-0.464} $ && $ 6.57^{+0.25}_{-0.24} $ & $ 0.501^{+0.072}_{-0.070} $ & $ 2.444^{+0.031}_{-0.030} $ & $ 1.649^{+0.575}_{-0.575} $ && $ 7.08^{+0.27}_{-0.26} $ & $ 0.516^{+0.078}_{-0.076} $ & $ 2.420^{+0.032}_{-0.030} $ & $ 1.469^{+0.513}_{-0.513} $  \\
A1835              & $ 8.54^{+0.08}_{-0.08} $ & $ 0.455^{+0.017}_{-0.016} $ & $ 2.940^{+0.007}_{-0.007} $ & $ 0.695^{+0.168}_{-0.168} $ && $ 7.14^{+0.05}_{-0.05} $ & $ 0.444^{+0.015}_{-0.014} $ & $ 2.948^{+0.007}_{-0.007} $ & $ 0.931^{+0.225}_{-0.225} $ && $ 7.60^{+0.07}_{-0.07} $ & $ 0.451^{+0.015}_{-0.015} $ & $ 2.904^{+0.007}_{-0.007} $ & $ 0.847^{+0.205}_{-0.205} $  \\
MACSJ1423.8$+$2404   & $ 6.25^{+0.15}_{-0.15} $ & $ 0.578^{+0.048}_{-0.045} $ & $ 2.269^{+0.022}_{-0.019} $ & $ 2.243^{+0.841}_{-0.841} $ && $ 5.77^{+0.12}_{-0.11} $ & $ 0.523^{+0.044}_{-0.042} $ & $ 2.251^{+0.021}_{-0.019} $ & $ 2.521^{+0.944}_{-0.944} $ && $ 6.28^{+0.15}_{-0.15} $ & $ 0.549^{+0.046}_{-0.044} $ & $ 2.233^{+0.024}_{-0.024} $ & $ 2.254^{+0.845}_{-0.845} $  \\
A1914              & $11.36^{+0.36}_{-0.36} $ & $ 0.364^{+0.065}_{-0.063} $ & $ 3.116^{+0.027}_{-0.026} $ & $ 0.596^{+0.158}_{-0.158} $ && $ 9.47^{+0.29}_{-0.24} $ & $ 0.372^{+0.052}_{-0.051} $ & $ 3.182^{+0.023}_{-0.023} $ & $ 0.812^{+0.214}_{-0.214} $ && $10.50^{+0.33}_{-0.34} $ & $ 0.386^{+0.058}_{-0.057} $ & $ 3.112^{+0.024}_{-0.024} $ & $ 0.678^{+0.180}_{-0.180} $  \\
A1995              & $ 9.09^{+0.30}_{-0.30} $ & $ 0.476^{+0.066}_{-0.064} $ & $ 2.606^{+0.025}_{-0.024} $ & $ 1.034^{+0.246}_{-0.246} $ && $ 7.78^{+0.25}_{-0.23} $ & $ 0.422^{+0.051}_{-0.049} $ & $ 2.610^{+0.022}_{-0.021} $ & $ 1.344^{+0.319}_{-0.319} $ && $ 8.40^{+0.27}_{-0.28} $ & $ 0.434^{+0.057}_{-0.054} $ & $ 2.579^{+0.023}_{-0.021} $ & $ 1.186^{+0.282}_{-0.282} $  \\
A2111              & $ 7.69^{+0.94}_{-0.78} $ & $ 0.364^{+0.219}_{-0.194} $ & $ 2.813^{+0.097}_{-0.085} $ & $ 0.903^{+0.440}_{-0.440} $ && $ 7.31^{+0.91}_{-0.73} $ & $ 0.169^{+0.177}_{-0.155} $ & $ 2.773^{+0.079}_{-0.060} $ & $ 0.921^{+0.449}_{-0.449} $ && $ 7.47^{+0.96}_{-0.77} $ & $ 0.170^{+0.181}_{-0.156} $ & $ 2.750^{+0.080}_{-0.059} $ & $ 0.893^{+0.437}_{-0.437} $  \\
A2163\tablenotemark{b}& $20.98^{+0.70}_{-0.69} $ & $ 0.277^{+0.065}_{-0.065} $ & $ 2.644^{+0.022}_{-0.023} $ & $ 0.247^{+0.065}_{-0.065} $ && $14.52^{+0.34}_{-0.34} $ & $ 0.344^{+0.040}_{-0.041} $ & $ 2.782^{+0.015}_{-0.015} $ & $ 0.485^{+0.127}_{-0.127} $ && $16.72^{+0.47}_{-0.48} $ & $ 0.353^{+0.048}_{-0.048} $ & $ 2.699^{+0.017}_{-0.017} $ & $ 0.372^{+0.097}_{-0.097} $  \\
A2163\tablenotemark{c}& $12.38^{+0.33}_{-0.31} $ & $ 0.319^{+0.037}_{-0.037} $ & $ 2.545^{+0.013}_{-0.012} $ & $ 0.641^{+0.168}_{-0.168} $ && $10.24^{+0.19}_{-0.18} $ & $ 0.318^{+0.028}_{-0.027} $ & $ 2.597^{+0.010}_{-0.010} $ & $ 0.884^{+0.230}_{-0.230} $ && $11.21^{+0.19}_{-0.19} $ & $ 0.334^{+0.031}_{-0.030} $ & $ 2.543^{+0.011}_{-0.010} $ & $ 0.755^{+0.196}_{-0.196} $  \\
A2204              & $ 8.06^{+0.09}_{-0.09} $ & $ 0.578^{+0.022}_{-0.021} $ & $ 3.034^{+0.010}_{-0.010} $ & $ 1.224^{+0.333}_{-0.333} $ && $ 6.49^{+0.06}_{-0.05} $ & $ 0.556^{+0.019}_{-0.018} $ & $ 3.042^{+0.009}_{-0.009} $ & $ 1.761^{+0.479}_{-0.479} $ && $ 6.92^{+0.06}_{-0.06} $ & $ 0.562^{+0.020}_{-0.020} $ & $ 2.986^{+0.009}_{-0.009} $ & $ 1.592^{+0.433}_{-0.433} $  \\
A2218              & $ 7.23^{+0.13}_{-0.18} $ & $ 0.269^{+0.056}_{-0.054} $ & $ 2.875^{+0.026}_{-0.024} $ & $ 1.524^{+0.405}_{-0.405} $ && $ 6.55^{+0.17}_{-0.17} $ & $ 0.263^{+0.046}_{-0.046} $ & $ 2.898^{+0.024}_{-0.024} $ & $ 1.747^{+0.466}_{-0.466} $ && $ 6.92^{+0.17}_{-0.17} $ & $ 0.246^{+0.051}_{-0.049} $ & $ 2.855^{+0.024}_{-0.024} $ & $ 1.596^{+0.426}_{-0.426} $  \\
RXJ1716.4+6708     & $ 6.29^{+0.78}_{-0.66} $ & $ 0.867^{+0.319}_{-0.263} $ & $ 2.215^{+0.127}_{-0.108} $ & $ 2.222^{+1.134}_{-1.134} $ && $ 5.96^{+0.76}_{-0.62} $ & $ 0.748^{+0.278}_{-0.229} $ & $ 2.175^{+0.118}_{-0.103} $ & $ 2.276^{+1.163}_{-1.163} $ && $ 6.10^{+0.79}_{-0.66} $ & $ 0.773^{+0.275}_{-0.239} $ & $ 2.176^{+0.119}_{-0.105} $ & $ 2.214^{+1.135}_{-1.135} $  \\
A2259              & $ 5.35^{+0.28}_{-0.27} $ & $ 0.400^{+0.129}_{-0.116} $ & $ 2.992^{+0.084}_{-0.069} $ & $ 0.629^{+0.457}_{-0.457} $ && $ 4.88^{+0.26}_{-0.24} $ & $ 0.325^{+0.113}_{-0.104} $ & $ 2.957^{+0.078}_{-0.069} $ & $ 0.698^{+0.508}_{-0.508} $ && $ 5.26^{+0.28}_{-0.27} $ & $ 0.364^{+0.117}_{-0.117} $ & $ 2.919^{+0.074}_{-0.070} $ & $ 0.634^{+0.461}_{-0.461} $  \\
A2261              & $ 8.36^{+0.28}_{-0.28} $ & $ 0.514^{+0.065}_{-0.063} $ & $ 3.022^{+0.029}_{-0.028} $ & $ 0.765^{+0.227}_{-0.227} $ && $ 7.33^{+0.21}_{-0.19} $ & $ 0.519^{+0.058}_{-0.057} $ & $ 3.059^{+0.028}_{-0.027} $ & $ 0.938^{+0.278}_{-0.278} $ && $ 7.90^{+0.26}_{-0.25} $ & $ 0.534^{+0.060}_{-0.060} $ & $ 3.003^{+0.027}_{-0.027} $ & $ 0.835^{+0.248}_{-0.248} $  \\
MS2053.7$-$0449      & $ 5.97^{+0.95}_{-0.76} $ & $ 0.234^{+0.206}_{-0.178} $ & $ 2.221^{+0.106}_{-0.097} $ & $ 2.343^{+1.025}_{-1.025} $ && $ 5.50^{+0.91}_{-0.68} $ & $ 0.180^{+0.210}_{-0.163} $ & $ 2.193^{+0.102}_{-0.103} $ & $ 2.547^{+1.116}_{-1.116} $ && $ 6.23^{+1.26}_{-0.92} $ & $ 0.190^{+0.217}_{-0.180} $ & $ 2.229^{+0.098}_{-0.098} $ & $ 2.099^{+0.965}_{-0.965} $  \\
MACSJ2129.4$-$0741   & $ 9.32^{+0.79}_{-0.67} $ & $ 0.646^{+0.145}_{-0.138} $ & $ 2.714^{+0.061}_{-0.056} $ & $ 1.105^{+0.356}_{-0.356} $ && $ 8.60^{+0.62}_{-0.59} $ & $ 0.487^{+0.123}_{-0.114} $ & $ 2.683^{+0.055}_{-0.052} $ & $ 1.197^{+0.381}_{-0.381} $ && $ 9.19^{+0.79}_{-0.66} $ & $ 0.514^{+0.134}_{-0.126} $ & $ 2.622^{+0.056}_{-0.052} $ & $ 1.094^{+0.353}_{-0.353} $  \\
RXJ2129.7$+$0005     & $ 5.56^{+0.23}_{-0.22} $ & $ 0.722^{+0.129}_{-0.118} $ & $ 2.910^{+0.071}_{-0.067} $ & $ 1.025^{+0.376}_{-0.376} $ && $ 5.36^{+0.22}_{-0.22} $ & $ 0.627^{+0.114}_{-0.105} $ & $ 2.898^{+0.067}_{-0.068} $ & $ 1.029^{+0.377}_{-0.377} $ && $ 5.55^{+0.23}_{-0.22} $ & $ 0.663^{+0.115}_{-0.106} $ & $ 2.878^{+0.064}_{-0.066} $ & $ 0.988^{+0.362}_{-0.362} $  \\
MACSJ2214.9$-$1359   & $10.18^{+0.81}_{-0.73} $ & $ 0.422^{+0.122}_{-0.123} $ & $ 2.794^{+0.050}_{-0.048} $ & $ 1.933^{+0.541}_{-0.541} $ && $ 8.16^{+0.53}_{-0.49} $ & $ 0.388^{+0.092}_{-0.090} $ & $ 2.771^{+0.044}_{-0.041} $ & $ 2.758^{+0.753}_{-0.753} $ && $ 8.81^{+0.64}_{-0.58} $ & $ 0.418^{+0.106}_{-0.104} $ & $ 2.714^{+0.046}_{-0.043} $ & $ 2.469^{+0.682}_{-0.682} $  \\
MACSJ2228.5$+$2036   & $ 8.40^{+0.69}_{-0.63} $ & $ 0.552^{+0.145}_{-0.134} $ & $ 2.816^{+0.065}_{-0.060} $ & $ 2.552^{+0.762}_{-0.762} $ && $ 7.49^{+0.60}_{-0.50} $ & $ 0.449^{+0.124}_{-0.115} $ & $ 2.792^{+0.060}_{-0.055} $ & $ 2.966^{+0.877}_{-0.877} $ && $ 8.33^{+0.67}_{-0.61} $ & $ 0.467^{+0.133}_{-0.125} $ & $ 2.742^{+0.060}_{-0.055} $ & $ 2.504^{+0.745}_{-0.745} $  \\
\enddata
\tablenotetext{a}{Units are $\times 10^{-15}$ counts s$^{-1}$ cm$^{3}$.}
\tablenotetext{b}{Using $N_{\mathrm{H}} = 12.1 \times 10^{20}$ cm$^{-2}$.}
\tablenotetext{c}{Using $N_{\mathrm{H}} = 18.7 \times 10^{20}$ cm$^{-2}$.}
\end{deluxetable*}

  \clearpage
\end{landscape}

\end{document}